\documentclass[%
 reprint,
 amsmath,amssymb,
 aps,
pra,
]{revtex4-2}

\usepackage{braket}
\usepackage{subcaption}
\usepackage{dsfont}
\usepackage[mark= ]{sectionbreak}

\usepackage{graphicx}
\graphicspath{{images/}}
\usepackage{dcolumn}
\usepackage{bm}
\usepackage{mathtools}
\usepackage[thinc]{esdiff}
\usepackage{textcomp}
\usepackage{xcolor}
\usepackage{caption}
\begin{document}

\title{ Robustness of diabatic enhancement in quantum annealing }

\author{Natasha Feinstein}
\email{ucapnjs@ucl.ac.uk}
\address{%
 London Centre for Nanotechnology, University College London, WC1H 0AH London, UK
}%

\author{Ivan Shalashilin}
\address{%
 Department of Physics, Imperial College London, SW7 2BW London, UK
}%

\author{Sougato Bose}
\address{%
 Department of Physics and Astronomy, University College London, London WC1E 6BT, UK
}%

\author{P. A. Warburton}
\address{%
 London Centre for Nanotechnology, University College London, WC1H 0AH London, UK
}%
\address{%
 Department of Electronic \& Electrical Engineering, University College London, WC1E 7JE London, UK
}%

\date{\today}


\begin{abstract}
In adiabatic quantum annealing, the speed with which an anneal can be run, while still achieving a high final ground state fidelity, is dictated by the size of the minimum gap that appears between the ground and first excited state in the annealing spectrum. To avoid the exponential slowdown associated with exponentially closing gaps, diabatic transitions to higher energy levels may be exploited in such a way that the system returns to the ground state before the end of the anneal. In certain cases, this is facilitated by the original annealing spectrum. However, there are also examples where careful manipulation of the annealing Hamiltonian has been used to alter the spectrum to create a diabatic path to the ground state. Since diabatic transitions depend on the evolution rate and the gap sizes in the spectrum, it is important to consider the sensitivity of any potential enhancement to changes in the anneal time as well as any parameters involved in the manipulation of the spectrum. We explore this sensitivity using annealing spectra containing an exponentially closing gap and an additional, tuneable, small gap created by a catalyst. We find that there is a trade-off between the precision needed in the catalyst strength and the anneal time in order to maintain the enhancement to the final ground state fidelity.
\end{abstract}

\maketitle


\section{ Introduction \label{sec:introduction} }

Quantum annealing (QA) is a continuous time quantum algorithm for solving combinatorial optimisation problems \cite{Apolloni1989, Finnila1994, Kadowaki1998}. It works by initialising a quantum system in the ground state (GS) of a simple driver Hamiltonian, conventionally a homogeneous transverse field, before evolving to a Hamiltonian whose GS encodes the optimal solution of the problem to be solved. If the evolution proceeds adiabatically the system will be measured to be in the GS at the end of the anneal. How quickly the interpolation can be carried out while keeping the system in the GS is determined by the adiabatic theorem \cite{TosioKato1950} - the simplest version of which states that the total evolution time must scale inversely with the square of the minimum gap encountered between the ground and first excited state (1ES) over the course of the evolution.

It was proposed that QA would be able to find the solution to combinatorial optimisation problems faster than classical algorithms by virtue of being able to tunnel out of local minima to find the global optimum \cite{Apolloni1989, Finnila1994, Kadowaki1998}. Early numerical results for small system sizes were promising, with randomly generated instances of the exact-cover problem having a minimum energy gap that closed polynomially \cite{Farhi2001} with the system size - which would allow QA to find the solution efficiently. However, subsequent numerical and analytical studies have been less promising. Crucially, the time required for the the system to tunnel between competing local optima has been found in general to be exponential in the problem size \cite{smelyanskiy2002simulations,Young2010,Hen_2011} - or, equivalently, for the minimum spectral gap between the GS and 1ES to close exponentially. In particular, this scaling has been found to be associated with the problem instances that are hardest to solve classically \cite{Young2010}. 

A particular problem is the appearance of perturbative crossings which can form as a result of a single local optimum that sits in a wider potential well than the global optimum or as a result of many local optima that are close in Hamming distance \cite{Choi2020,Choi2021,Altshuler2010,Amin2009}. These perturbative crossings result in the formation of a gap minimum towards the end of the anneal which is exponentially small in the Hamming distance between the local and global optima \cite{Amin2008,Amin2009,Altshuler2010}. This is generally expected to grow linearly with the problem size. In addition, exponentially closing gaps have been observed even without the problem of competing local optima as a result of first order phase transitions between the delocalised initial GS and the localised GS of the problem Hamiltonian \cite{Jorg2010,Jorg2010a,Seoane2012,Knysh2016}.

These results confirm the standard assumption that quantum algorithms such as QA will not be able to efficiently solve hard instances of NP-problems \cite{Preskill_2018}. However, even if QA cannot change the complexity class of the problem, the algorithm may still offer a quantitative improvement to the run-time scaling. Indeed, such improvements have been found with QA showing a polynomial improvement in time scaling over simulated annealing for finding the GS of random Ising chains \cite{Zanca2016} as well as solving the Max Independent Set (MIS) problem \cite{Ebadi2022}. Further, the existence of a Grover-like quadratic speedup was proved for QA applied to unstructured search for some interpolation between $H_d$ and $H_p$ \cite{Roland_2002} - though it may be the case that the schedule that yields this speedup is not practical to find \cite{villanueva2022adiabatic}. 

However, even if we are satisfied with a polynomial reduction to the run-time, exponentially long annealing runs still pose a problem as they demand exponentially long coherence times from the hardware. While coherent quantum evolution has been demonstrated on the D-Wave Advantage \cite{king2022}, significant effects of decoherence were seen for anneal times greater than a few tens of nanoseconds. Other platforms may have longer coherence times than this \cite{Ebadi2022}. However, without the error correcting framework that exists for gate based quantum computation, exponentially long annealing runs remain intractable. A potential solution is to weaken the condition that the evolution must remain adiabatic. For instance, rather than running the anneal slowly enough to obtain a high final GS overlap, a fast anneal, which we might expect to have an exponentially low overlap with the final GS, could be repeated an exponential number of times such that the probability of obtaining the optimal solution is still high. The total time to solution (TTS) still scales exponentially, however the time for which an individual anneal is run does not. 

In this situation, the diabaticity is not helping the system reach the GS. Rather, it is an unwanted side effect of running the anneal within the coherence time of the hardware for which we compensate with repeated runs. However, there are also examples where allowing the system to evolve diabatically is useful in helping the system reach a higher overlap with the GS. For the glued trees problem, the structure of the annealing spectrum allows the system to reach the GS of the problem Hamiltonian in polynomial time by utilising diabatic transitions between energy levels \cite{Somma2012}. While this result is dependent on the highly symmetric nature of the problem, it serves as a proof of concept that a suitable annealing spectrum can facilitate a diabatic route to the GS for polynomial anneal times. More generally, it has been noted in multiple settings \cite{Crosson2014,Wecker_2016,Zhou_2020,Muthukrishnan2016} that, when restricted to modest anneal times, faster anneals actually achieved a greater overlap with the GS at the end of the anneal. The authors account for this as follows: if there is a small gap towards the end of the anneal then it may be beneficial for some of the amplitude to leak out of the GS prior to this point. That way some of the amplitude may be returned to the GS at the location of the gap minimum - rather than the evolution being totally adiabatic until the small gap at which point all the GS amplitude may be lost. 

The run-time, however, may need to chosen very precisely in order for diabatic quantum annealing (DQA) to maintain its advantage over adiabatic QA \cite{Muthukrishnan2016,Brady_2017}. This is because in DQA the final GS overlap is a result of a complex interplay of transitions which depend on the speed at which the anneal is run. In order to understand the practicality of using DQA strategies, it will be crucial to understand how the requirements on the precision in annealing time relate to the precision with which we can expect to know the optimal annealing time \textit{a priori} - as well as any hardware considerations on setting it. 

Further to this, more recent work has looked at manipulating the annealing spectrum to create a diabatic path to the GS. Two ways this can be done are through the addition of a catalyst Hamiltonian \cite{Choi2021, feinstein2023effects} or by inhomogeneous driving \cite{Fry-Bouriaux2021}. Strategies such as these are appealing since they could enable us to produce spectra that allow the final GS to be reached with polynomial anneal times - rather than having to rely on the problem naturally producing such a spectrum. In these cases it will be important to understand the robustness of the success of these approaches to changes in not only the anneal time, but also the parameters used to manipulate the spectrum.

In this work, we gain insight into the robustness of DQA strategies in which the spectrum is altered to create a diabatic path to the GS. Our findings suggest a trade-off between the precision needed in the parameters used to alter the annealing spectrum and the anneal time chosen - with greater precision in one resulting a greater robustness to imprecision in the other. We utilise a toy model from our previous work \cite{feinstein2023effects}. We believe however that these findings should apply to other settings where the spectrum is manipulated in similar ways.

We start by outlining our problem setting and the catalyst Hamiltonian we use to alter the annealing spectrum in section \ref{sec:setting}. In section \ref{sec:fidelity_enhancement}, we show that this catalyst has the capability to guide the system to a final GS overlap of unity for significantly faster anneal times than are needed in the catalyst-free case. In section \ref{sec:robustness} we present numerical results for the final GS fidelity with varying catalyst strengths and anneal times - where we observe the aforementioned trade-off between the precision needed in the parameters. We then shed some light on these results through the theory of Landau-Zener transitions in section \ref{sec:LZ} before discussing the implications of our results in section \ref{sec:discussion}.

\section{ Problem setting \label{sec:setting} }

In this section we describe our problem setting and demonstrate the effect the catalyst has on the annealing spectrum. 

We have made use of the maximum weighted independent set (MWIS) problem to create problem instances that result in the appearance of a perturbative crossing in the annealing spectrum for a standard linear anneal. Perturbative crossings are a particular source of exponentially closing gaps which can form as a result of highly competitive local optima that sit in wider potential wells than the global optimum. These can be the result of many local optima being close in Hamming distance to each other or simply a single local optimum surrounded by many low energy states. We outline the theoretical background to the formation of perturbative crossings in appendix \ref{app:PCs}.

Specifically, the instances of the MWIS problem that we create have a single local optimum in addition to the global optimum resulting in a single perturbative crossing between the GS and 1ES towards the end of the anneal. We utilised this setting in a previous work \cite{feinstein2023effects} to investigate the effect of a targeted XX-catalyst on the annealing spectrum and how sensitive this effect was to small changes to the problem Hamiltonian. Among other results, we found that under certain conditions, the introduction of our chosen catalyst resulted in the appearance of an additional small gap between the GS and 1ES. The size this small gap can be tuned by adjusting the strength with which the catalyst was introduced. This additional gap minimum produced a diabatic path which could be utilised to achieve a final GS fidelity close to unity for significantly shorter run-times.

We begin this section by describing our problem setting in section \ref{sec:nocat}. We outline how the graph structure results in a perturbative crossing and present numerical results showing the existence of an exponentially closing gap. In \ref{sec:withcat}, we discuss the effects of the catalyst. We will briefly discuss our choice of catalyst, present numerical results showing that its introduction creates a new gap minimum in the spectrum, and characterise the formation of this new gap minimum with increasing system size.

\subsection{ Catalyst-free setting \label{sec:nocat} }

\begin{figure*}[]
    \centering
    \includegraphics[scale=0.37]{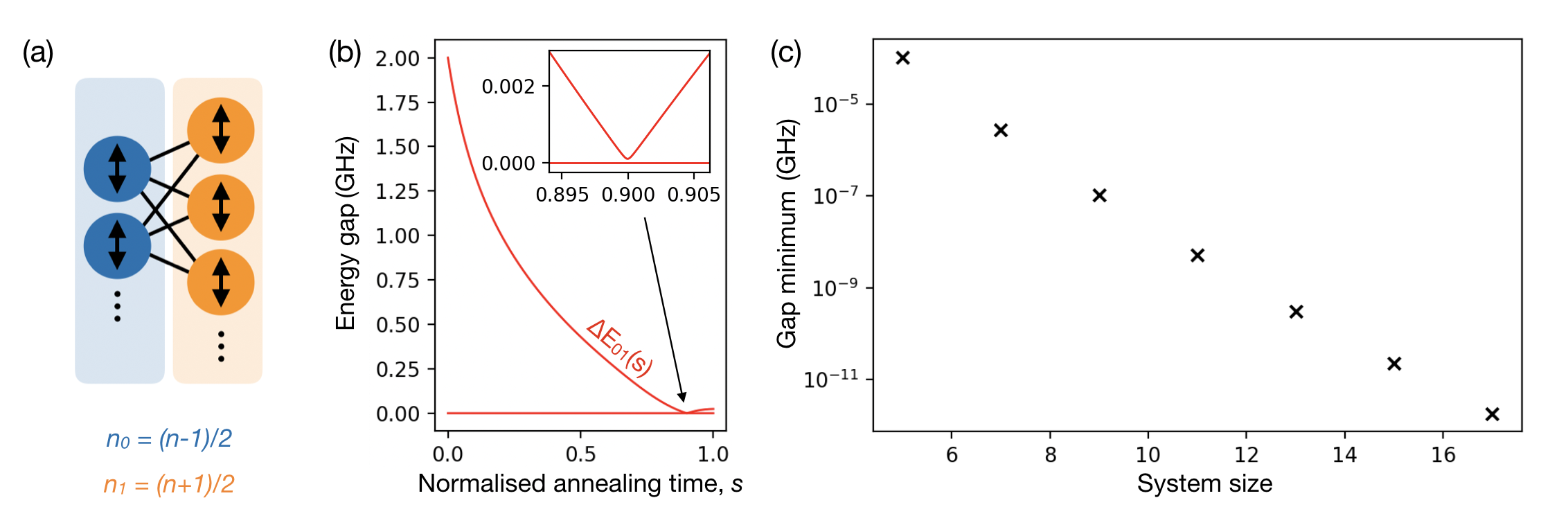}
    \caption{ An illustration of our problem graph structure is shown in (a). Plots (b) and (c) show numerical results for the annealing spectrum in the catalyst-free setting. The spectrum for the 5-vertex instance is shown in (b) and the scaling with system size of the minimum gap in the spectrum is shown in (c). }
    \label{fig:catalyst_free}
\end{figure*}

In the catalyst-free case, the annealing Hamiltonian is given by
\begin{equation}
    H(s) = (1-s) H_d + s H_p,
    \label{eq:H(s)}
\end{equation}
where $H_d$ and $H_p$ are independent of $s$ and denote the driver and problem Hamiltonians respectively. The dimensionless annealing parameter, $s$, is varied from $0$ to $1$ over the course of the anneal such that $H(s)$ evolves from $H(0)=H_d$ to $H(1)=H_p$. Specifically, we take $s=t/t_a$ where $t_a$ is the total anneal time. 

We refer to the eigenstates of the total Hamiltonian as the \textit{instantaneous} eigenstates and denote them and their corresponding energies as
\begin{equation}
    H(s) \ket{E_a(s)} = E_a(s) \ket{E_a(s)}.
    \label{eq:insts}
\end{equation}
The states are labelled starting from $a=0$ in order of increasing energy. Similarly we denote the \textit{problem} eigenstates and their energies as
\begin{equation}
    H_p \ket{E_a} = E_a \ket{E_a}.
    \label{eq:probs}
\end{equation}
Since the problem Hamiltonian is diagonal in the computational basis, the set of problem states $\{\ket{E_a}\}$ is simply the computational basis with the states labelled by energy. Our choice of driver is the conventional homogeneous local X-field,
\begin{equation}
    H_d = - \sum \limits_{i = 1}^{n} \sigma_i^x,
    \label{eq:Hd}
\end{equation}
where $n$ is the total number of qubits and $\sigma_x^i$ denotes the Pauli-X operator on the $i$th qubit. Its GS, in which the system is initialised in at $s=0$, is the equal superposition over all computational basis states.

We now turn to the MWIS problem and its Ising encoding. The MWIS problem takes as its input an undirected weighted graph and its solution is the independent subset of vertices with the largest total weight. In its Ising formulation, each vertex of the problem graph is represented by a spin such that all possible sets of vertices can be represented by the basis states - with spin up denoting a vertex that is in the set and spin down denoting a vertex that is not.  For instance, the problem state $\ket{E_a} = \ket{\downarrow \uparrow \uparrow \downarrow \uparrow}$ encodes the set of vertices $\{2,3,5\}$. Note that for this encoding flipping a spin corresponds to either adding or removing a vertex from the set. Local Z-fields are used to set the vertex weights and the independent set condition is implemented by introducing an anti-ferromagnetic ZZ-coupling between any two qubits corresponding to vertices connected by an edge. Overall, the problem Hamiltonian is given by
\begin{equation}
\label{eq:Problem-Hamiltonian}
    H_p = \sum \limits_{i \in \{\textrm{\scriptsize vertices}\}} (c_i J_{zz} - 2w_i)\sigma^z_i + \sum \limits_{(i,j) \in \{\textrm{\scriptsize edges}\}} J_{zz} \sigma^z_i \sigma^z_j
\end{equation}
where $c_i$ is the degree and $w_i$ the weight of vertex $i$. $J_{zz}$ can be tuned to adjust the severity of the edge penalty implementing the independent set condition. That the edge penalty appears in the local field terms to account for the fact that two neighbouring down spins should not be penalised. This corresponds to two adjacent vertices \textit{not} being in the set which does not result in a dependency.

We now describe the MWIS instances that will appear in this work and how their structure results in the formation of a perturbative crossing. As described earlier, perturbative crossings can form as a result of highly competitive local optima. They can be understood through treating $H_d$ as a perturbation to $H_p$ and considering the energy corrections to the problem eigenstates \cite{Amin2009,Altshuler2010,Dickson2011a}. When the perturbations are such that the perturbed energy of one of the low lying excited states crosses that of the GS, we can expect an avoided level crossing to form towards the end of the annealing spectrum. Since the effect of the driver on the eigenstates of $H_p$ is to perform single spin flips, the perturbation to the eigenstates will depend on the states closest in Hamming distance to them. 

We refer the reader to appendix \ref{app:PCs} for an outline of the theory and to references \cite{Amin2009,Altshuler2010,Dickson2011a} for a more complete discussion on the formation of perturbative crossings. The key conclusion that we make use of in the construction of our problem instances is as follows: a perturbative crossing will form between the GS, $\ket{E_0}$, and some other state $\ket{E_a}$ that is close in energy, if $\ket{E_a}$ is coupled to more low energy states by the driver than $\ket{E_0}$ is. When this is the case, the instantaneous GS of $H(s)$ will first localise to $\ket{E_a}$ before sharply transitioning to $\ket{E_0}$ at the end of the anneal resulting in the system needing to tunnel from $\ket{E_a}$ to $\ket{E_0}$ at this point in order to end in the GS. That the system localises to a state surrounded by low energy states can be understood as the algorithm getting caught in a wide local optimum.

We now discuss how our problem instances are constructed. The local optima of the MWIS problem are the maximally independent subsets of the graph - that is, any independent set of vertices for which no other vertex can be added without including an edge. Thus, a straightforward way of creating an MWIS problem with two local optima is with a complete bipartite graph. Our problem setting consists of such MWIS instances. The local optima are then simply the selection of all the vertices in one of the two disjoint subgraphs, which we will denote $G_0$ and $G_1$ - see Figure \ref{fig:catalyst_free}(a). We then allocate a total weight, $W_0$ and $W_1$ respectively, to each subgraph which is split evenly between the vertices - \textit{i.e}: all the vertices in $G_i$ have weight $W_i/n_i$ where $n_i$ in the number of vertices in $G_i$. By selecting $W_0 > W_1$, $G_0$ will correspond to the global optimum and thus the GS of $H_p$, $\ket{E_0}$. By then selecting $W_1$ to be very close to $W_0$, $G_1$ becomes a highly competitive local optimum corresponding to the 1ES of $H_p$, $\ket{E_1}$.

To create a perturbative crossing between $\ket{E_0}$ and $\ket{E_1}$, we must ensure that $\ket{E_1}$ is coupled to more low energy states by the driver. This we achieve by giving $G_0$ fewer vertices than $G_1$. To understand how this works, recall that the action of $H_d$ on the problem eigenstates is to flip one of the spins. Also recall that, with respect to the MWIS problem, flipping one of the spins in $\ket{E_a}$ from down to up corresponds to adding a vertex to the corresponding set, and that flipping a spin from up to down corresponds to removing a vertex. Since $\ket{E_0}$ and $\ket{E_1}$ correspond to maximally independent sets, adding a vertex will result in a dependent set while removing one will of course result in another independent set. Thus, denoting the number of vertices in the two subgraphs as $n_0$ and $n_1$ respectively, $\ket{E_0}$ is coupled by $H_d$ to $n_0$ states corresponding to independent sets and $n_1$ states corresponding to dependent sets. The reverse is true of $\ket{E_1}$. Due to the edge penalty encoding the independent set condition, states corresponding to dependent sets will have significantly higher energy than those corresponding to independent sets. By choosing $n_1 > n_0$, we therefore create a setting where $\ket{E_1}$ is coupled to more low energy states than $\ket{E_0}$. Specifically, we take $n_0 = (n-1)/2$ and $n_1 = (n+1)/2$ (where $n$ is the total number of vertices) such that $G_1$ always has one more vertex than $G_0$. The specific parameter choices used in equation \ref{eq:Problem-Hamiltonian} for our problem instances are discussed in appendix \ref{app:prob_instances}.

In Figure \ref{fig:catalyst_free}(b) we show the gap spectrum corresponding to the $5$-vertex instance. These results are obtained by numerical diagonalisation of $H(s)$. The expected gap minimum can clearly be seen at $s \approx 0.9$. Figure \ref{fig:catalyst_free}(c) shows that the gap minimum closes exponentially as predicted by the theory (see appendix \ref{app:PCs}).

\subsection{ Introduction of XX-catalyst \label{sec:withcat} }

\begin{figure*}[]
    \centering
    \includegraphics[scale=0.35]{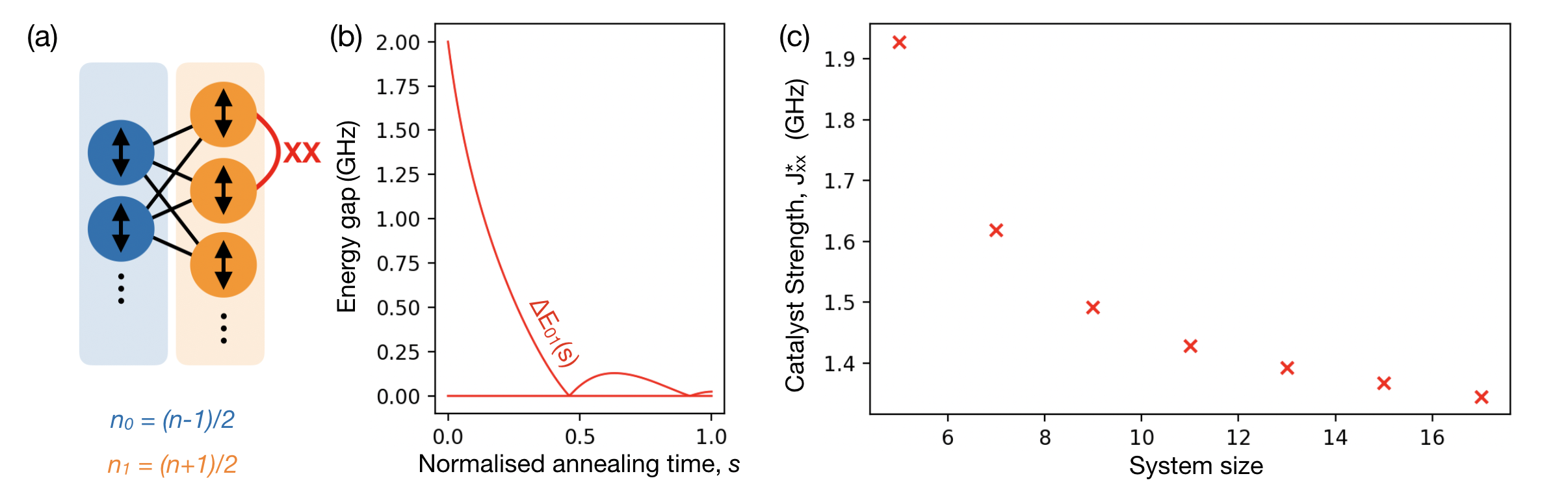}
    \caption{ An illustration of our graph structure is shown in (a) with the location of the XX-coupling indicated. Plots (b) and (c) show numerical results for the spectral properties when the catalyst is introduced. The gap spectrum of the $5$-vertex instance with the catalyst introduced with $J_\text{xx} = J^*_\text{xx}$ is shown in (b). The scaling of $J^*_\text{xx}$ with system size is shown in (c). }
    \label{fig:with_catalyst}
\end{figure*}

\begin{figure}[]
    \centering
    \includegraphics[scale=0.36]{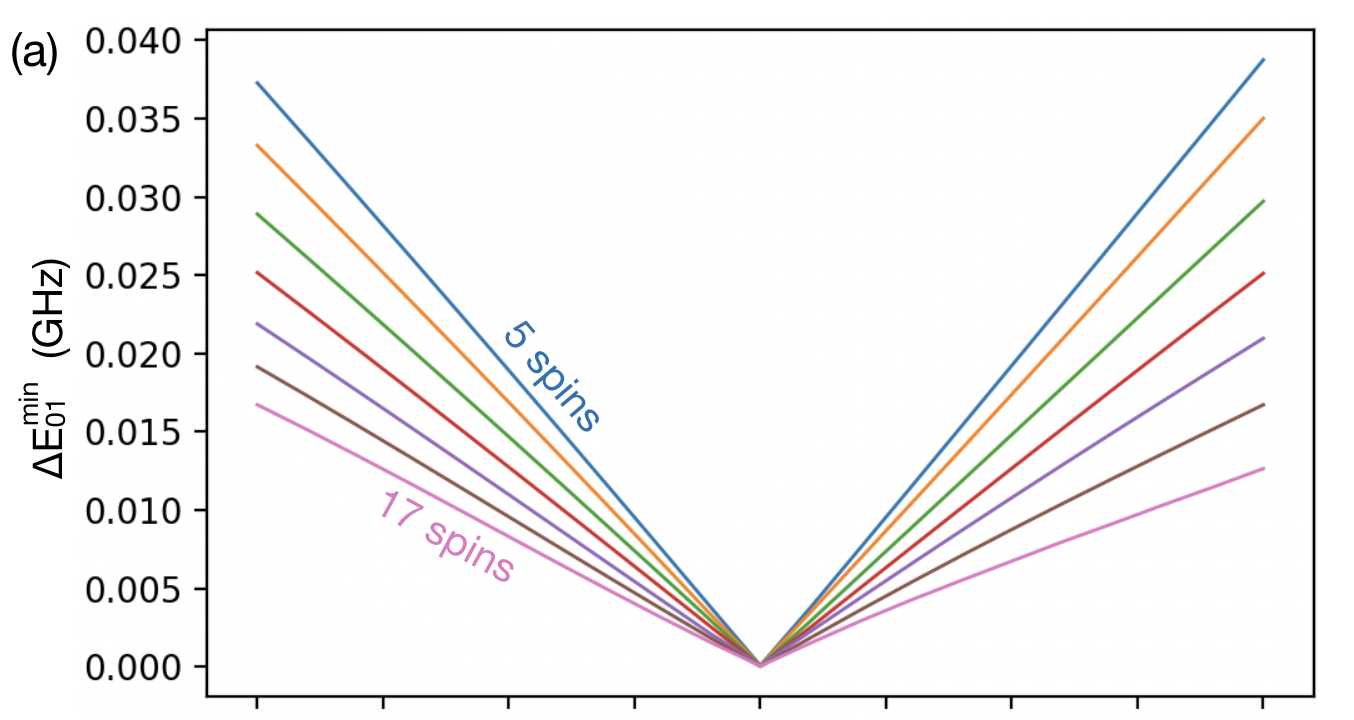}
    \includegraphics[scale=0.36]{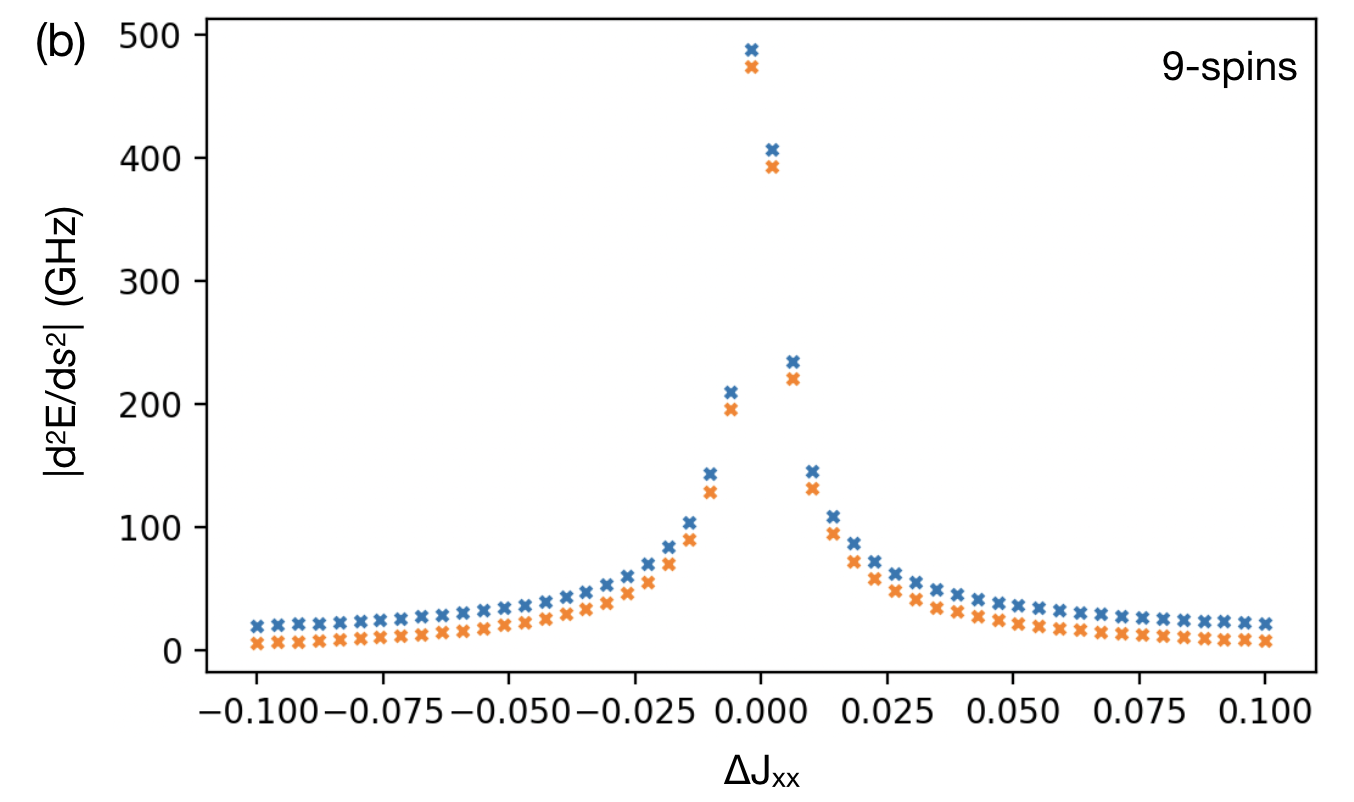}
    \caption{ In this figure we present properties of the gap spectra with varying $\Delta J_\text{xx}$. Plot (a) shows how the size of the new gap minimum varies for system sizes ranging from $5$ to $17$ in steps of $2$. Plot (b) shows the magnitude of the second derivative of the ground (blue) and first excited state (orange) energies at the gap minimum for the $9$-spin system. }
    \label{fig:spectrum_properties}
\end{figure}

We now introduce a catalyst such that the total Hamiltonian reads
\begin{equation}
    H(s) = (1-s)H_d + s(1-s)H_c + sH_p.
    \label{eq:Hc(s)}
\end{equation}
The catalyst, $H_c$, consists of a single XX-coupling between two vertices in $G_1$ as shown in Figure \ref{fig:with_catalyst}(a). By symmetry, all pairs of vertices in $G_1$ are equivalent. The catalyst Hamiltonian is thus written
\begin{equation}
    H_c = J_\text{xx} \sigma^x_i \sigma^x_j 
    \label{eq:Hc}
\end{equation}
where $i$ and $j$ are spins corresponding to vertices in $G_1$. We select $J_\text{xx}$ to be positive such that the XX-coupling enters $H(s)$ with the opposite sign to the driver terms resulting in a Hamiltonian that is non-stoquastic for $s \neq 0,1$. The magnitude of $J_\text{xx}$ can be tuned to adjust the strength of the catalyst relative to $H_d$ and $H_p$. There has been much discussion in the literature regarding the importance of non-stoquasticity in QA \cite{gupta2019elucidating,Choi2021,Aharonov2007}. We note that here we are using the term non-stoquastic simply to describe a Hamiltonian with positive as well as negative off diagonal elements and make no assumptions regarding the relationship between non-stoquasticity and computational complexity. 

For the motivation behind this catalyst we refer the reader to our previous work in which our initial investigations using this catalyst can be found \cite{feinstein2023effects}. In that study we found that this choice of catalyst created a new gap minimum in the spectrum when introduced with specific strengths, $J_\text{xx}$. This gap was found numerically to go to zero for some critical value of $J_\text{xx}$ which varied with system size. We will refer to this value as the optimal catalyst strength, $J^*_\text{xx}$, for reasons that will become clear in our results section. In Figure \ref{fig:with_catalyst}(b) we show the annealing spectrum for the $5$-vertex instance with $H_c$ introduced with the optimal catalyst strength for that system size. The perturbative crossing present in the original annealing spectrum can still be seen at $s \approx 0.9$. There is, however, now an additional gap minimum a little before $s=0.5$. The optimal catalyst strength is plotted against system size in Figure \ref{fig:with_catalyst}(c).

Decreasing or increasing $J_\text{xx}$ away from $J^*_\text{xx}$ results in an increase in the size of the new gap minimum produced by the catalyst. We define the quantity $\Delta J_\text{xx} = (J_\text{xx} - J^*_\text{xx})/J^*_\text{xx}$ and plot the gap dependence on $\Delta J_\text{xx}$ for different system sizes in Figure \ref{fig:spectrum_properties}(a). An approximately linear dependence on $|\Delta J_\text{xx}|$ is observed with the rate of increase differing slightly depending on whether $\Delta J_\text{xx}$ is positive or negative. This difference becomes more pronounced for larger system sizes. Note that the linear dependence breaks down for sufficiently large $|\Delta J_\text{xx}|$.

\section{ Results \label{sec:results} }

We now consider the system dynamics over the course of the anneal to (a) establish that the introduction of the new gap minimum does allow the system to reach the final GS for faster anneal times and (b) investigate how robust any enhancement to the GS fidelity is to changes in $J_\text{xx}$ and $t_a$. Our results are obtained using closed system spin models. To make the simulations more tractable we use the symmetric nature of the problem to reduce the Hilbert space - details of how this was done can be found in appendix \ref{app:hilbert-space}.

In section \ref{sec:fidelity_enhancement}, we show that the creation of the new gap minimum does allow the system to reach significantly higher final GS fidelities for shorter run-times. We also confirm that the enhancement is due to the expected diabatic transitions between the instantaneous GS and 1ES. The main results of this work are then presented in Sections \ref{sec:robustness} and \ref{sec:LZ}. Section \ref{sec:robustness} examines the robustness of the fidelity enhancement to changes in the catalyst strength and anneal time. Then, in section \ref{sec:LZ}, we use the Landau-Zener model for diabatic transitions to explain the physics behind our results.

Throughout this section, we supplement our discussion with numerical results for the 9-spin system. Similar results were also obtained for the other system sizes examined ($n=5$, $7$, $11$, $13$, $15$ and $17$).

\subsection{ Final GS fidelity enhancement \label{sec:fidelity_enhancement} }

\begin{figure}[]
    \centering
    \includegraphics[scale=0.29]{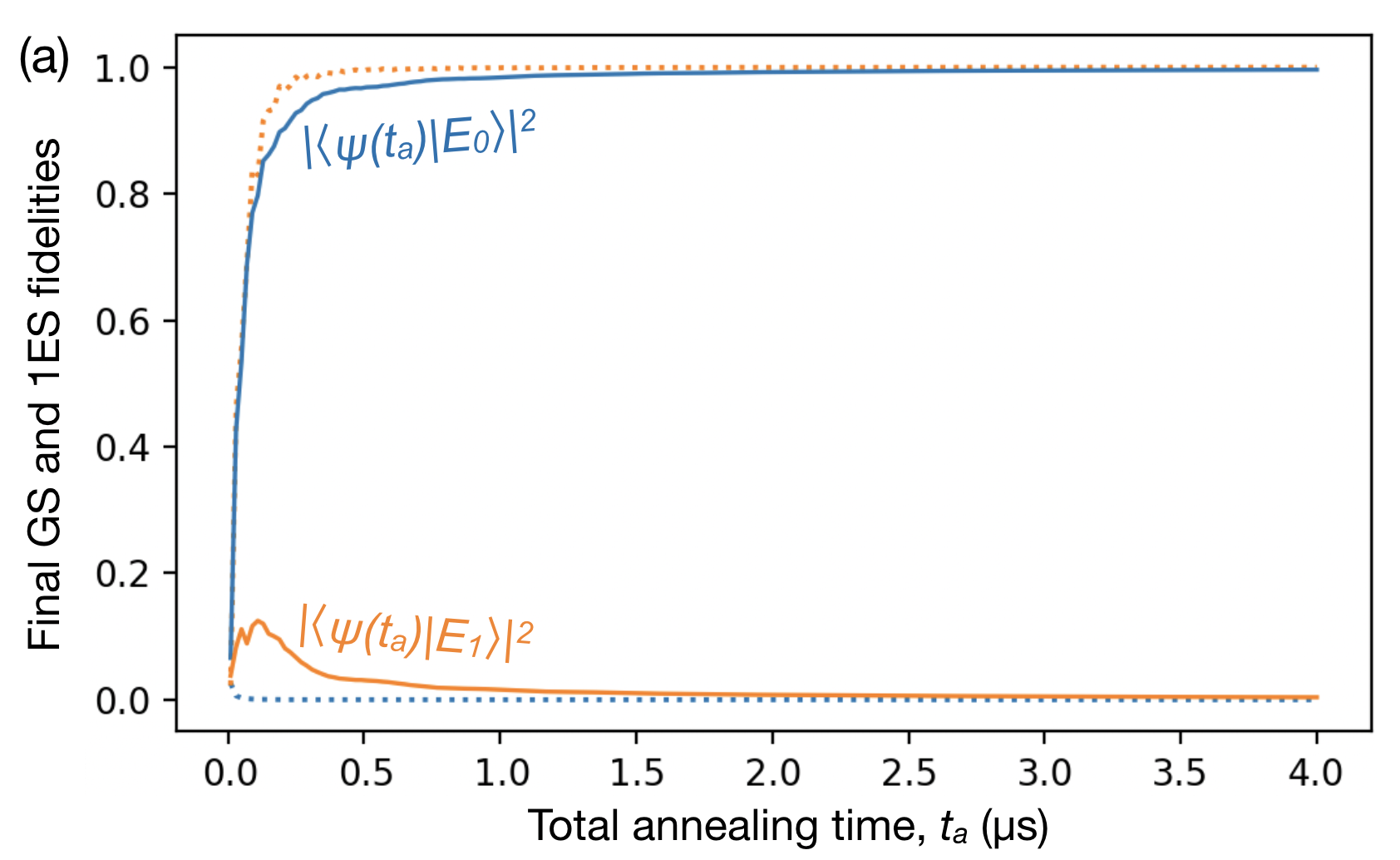}
    \includegraphics[scale=0.29]{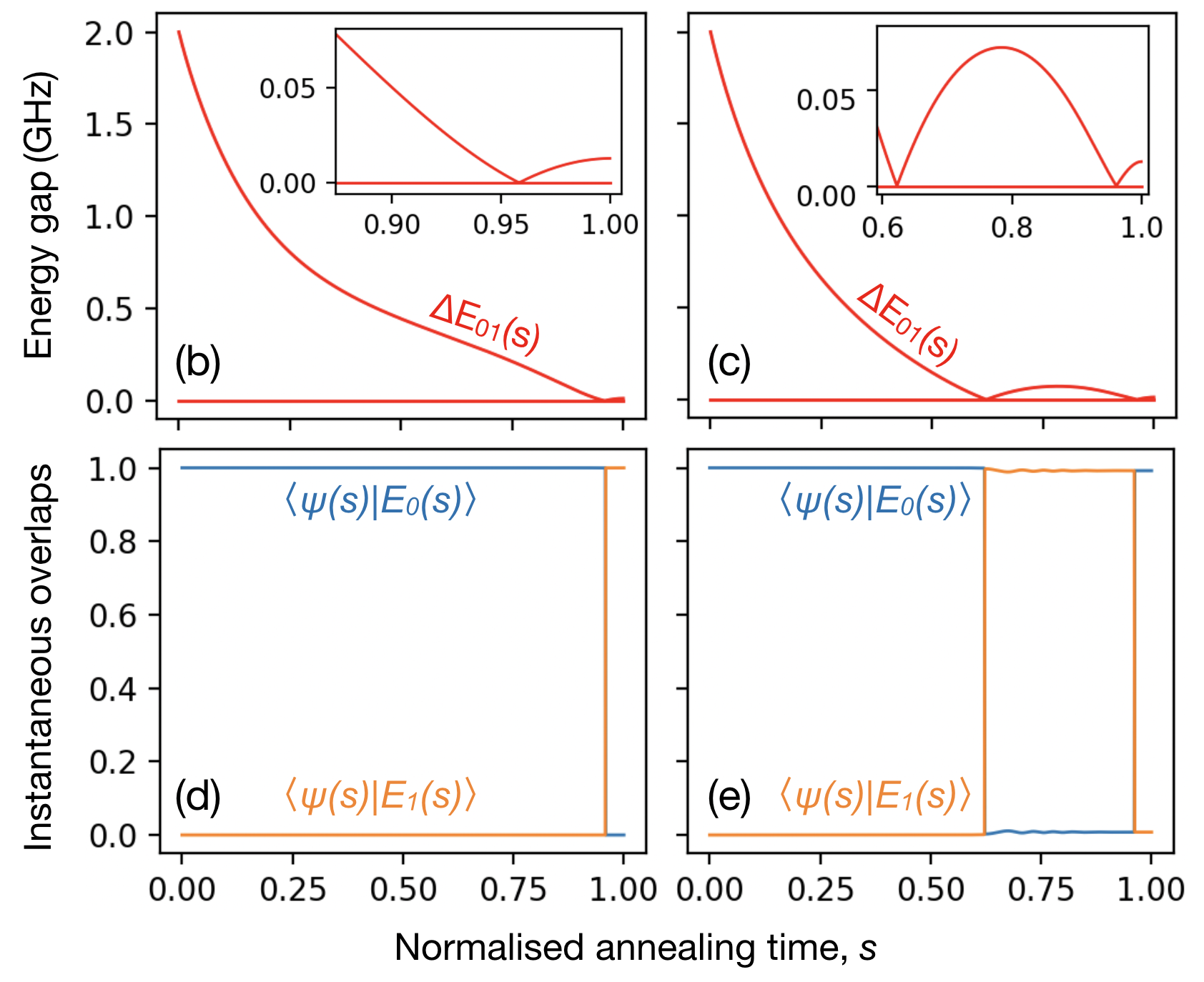}
    \caption{ Plot (a) shows the GS and 1ES fidelities at the end of the anneal ($t=t_a$) in blue and orange respectively. These results are for the 9-spin instance and were obtained using closed system spin models. The results without a catalyst are shown with dotted lines and the results with the catalyst introduced with $J_\text{xx} = J^*_\text{xx}$ are shown with solid lines. Plots (b) and (c) show the gap spectrum for the 9-vertex instance without and with a catalyst respectively. Plots (d) and (e) show the corresponding dynamics for a $t_a = 2\mu$s anneal with the state of the system represented in terms of its overlap with the instantaneous ground and first excited states in blue and orange respectively. }
    \label{fig:dynamics}
\end{figure}

Figure \ref{fig:dynamics}(a) shows the final GS (blue) and 1ES (orange) fidelities with increasing anneal time for the $9$-spin system. The results without the catalyst are shown in dotted lines and with the catalyst in solid lines. The local driver fields are introduced with a magnitude of $1$GHz, reflective of the energy scales in the D-Wave quantum annealer \cite{dwavedoc}. The magnitude of the catalyst is set to its optimum value, $J^*_\text{xx}$, for this system size. Without the catalyst, we see that the final GS fidelity is negligible for all the anneal times plotted and that the system ends in the 1ES. 

Looking at Figures \ref{fig:dynamics}(b) and (d), which show gap spectrum and dynamics at $t_a = 2\mu$s, we see that this is a result of the expected transition into the 1ES at the location of the gap minimum. We were not able to reach the adiabatic limit in our simulations. We however estimate the time needed to reach a near unity GS fidelity to be on the order of $10^4 \mu$s by extrapolating from simulations of shorter annealing runs.

From Figure \ref{fig:dynamics}(a) we see that, by introducing the catalyst with $J^*_\text{xx}$, the system is able to approach a final GS fidelity of unity for $t_a$ on the order of $1 \mu$s - an improvement of around four orders of magnitude. Looking at Figures \ref{fig:dynamics}(c) and (e), we can see that this is a result of the system transitioning to the 1ES at the new gap minimum created by the catalyst, and then back into the GS at the perturbative crossing.

\subsection{ Robustness of enhancement \label{sec:robustness} }

\begin{figure}[]
    \centering
    \includegraphics[scale=0.27]{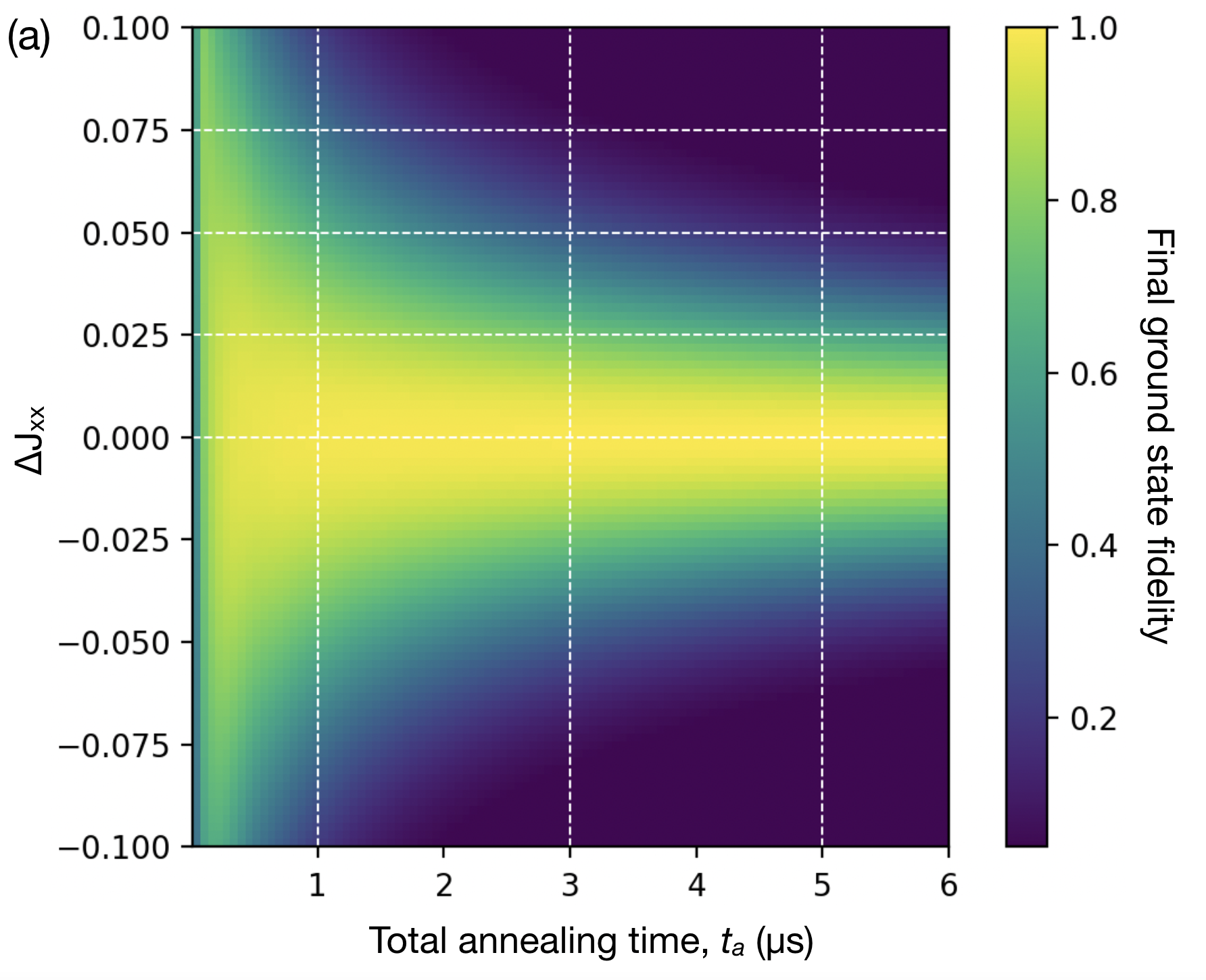}
    \includegraphics[scale=0.27]{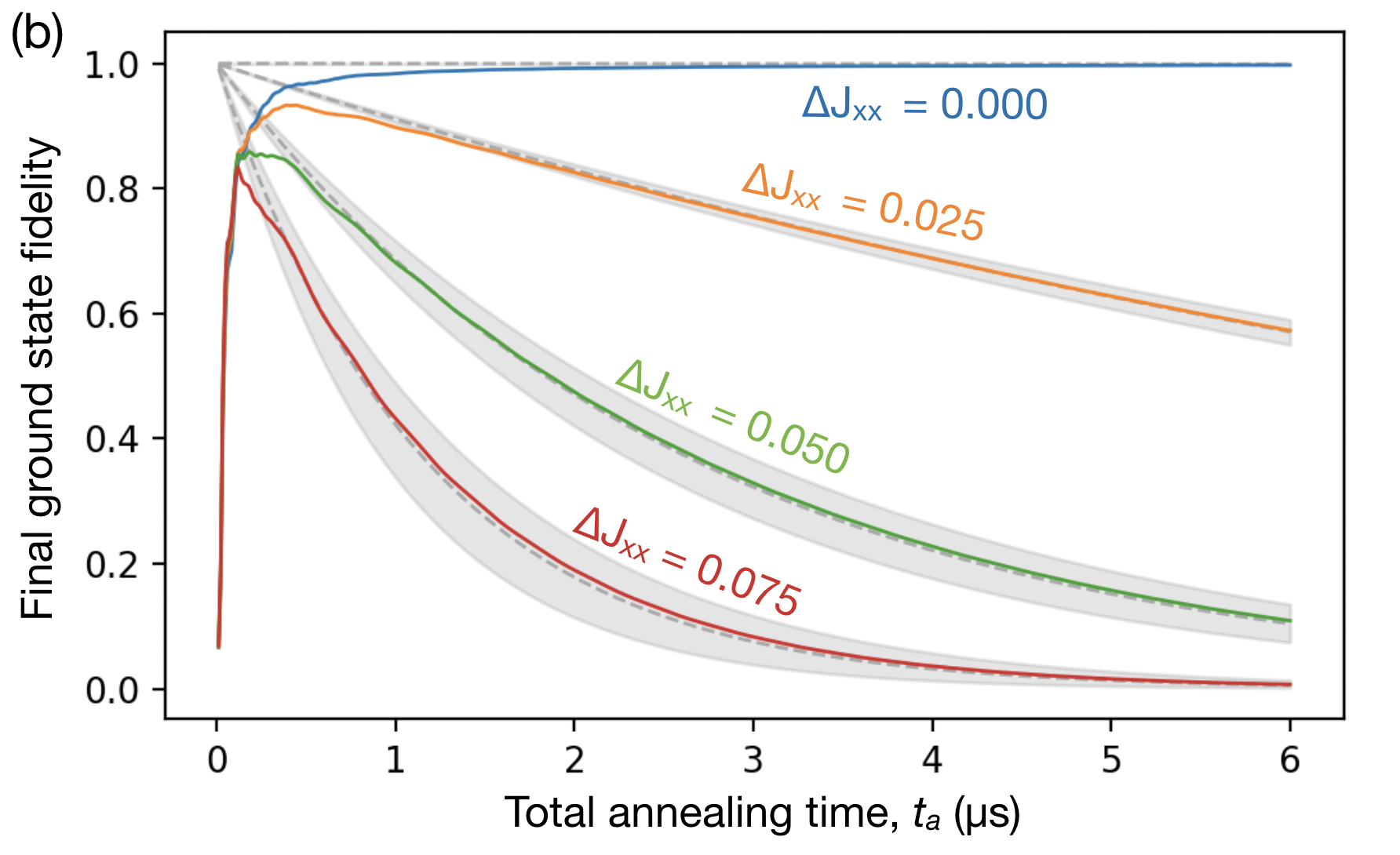}
    \includegraphics[scale=0.27]{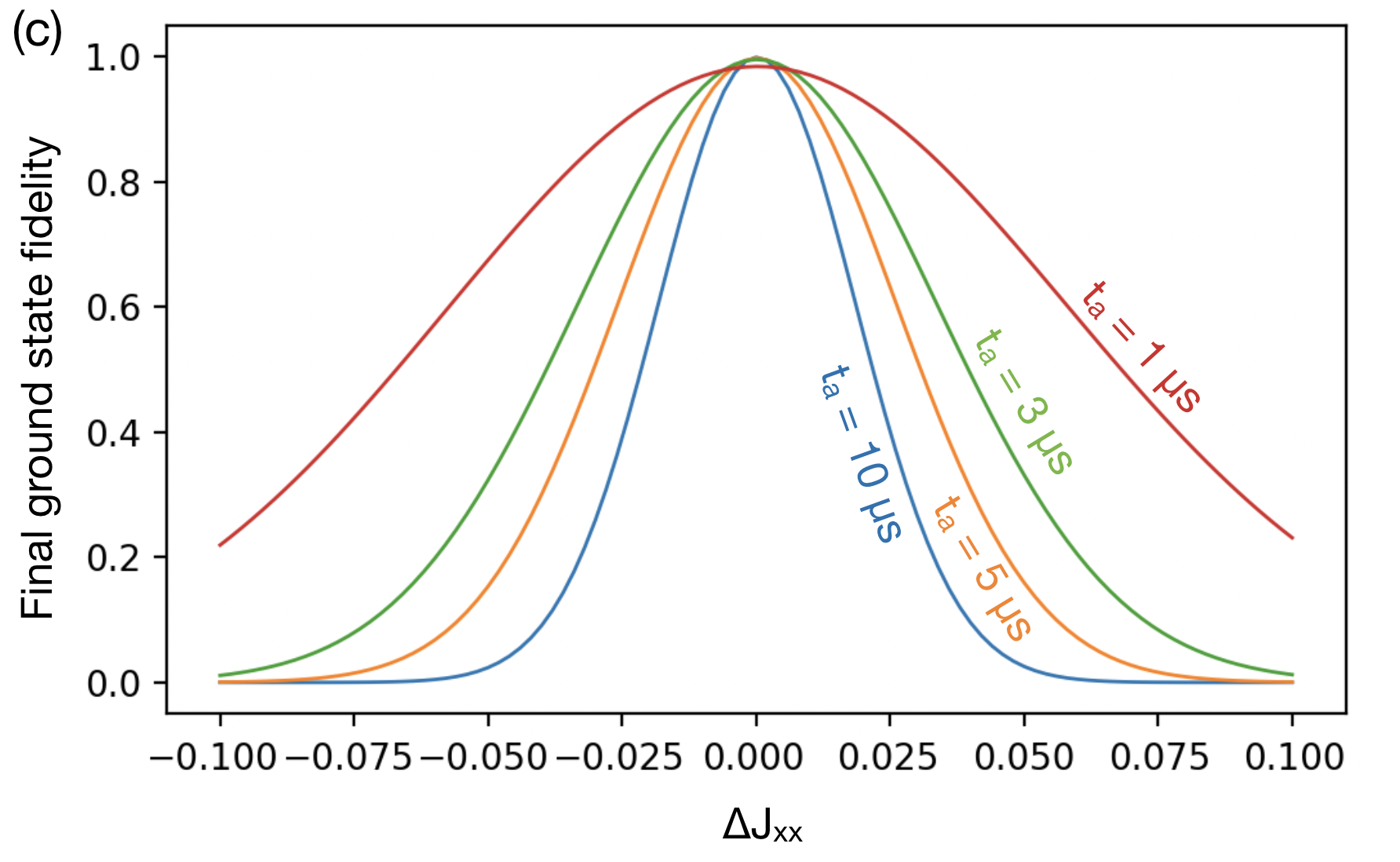}
    \caption{ This figure shows numerical results for the final GS fidelity with varying $t_a$ and $\Delta J_\text{xx}$ for the 9-spin system. Figure (a) shows a grid plot and Figures (b) and (c) show the slices of this grid indicated with dashed white lines. (b) shows the final GS fidelity with increasing anneal time, $t_a$, for different values of $\Delta J_\text{xx}$. The coloured curves show the numerical results and the grey shaded areas and dashed curves show the Landau-Zener predictions - obtained as described in Section \ref{sec:LZ}. (c) shows the results for the final GS fidelity with varying $\Delta J_\text{xx}$ for different values of $t_a$.}
    \label{fig:temp}
\end{figure}

Having established that the catalyst has the capacity to enhance the final GS fidelity, we now turn to the question of how robust this enhancement is. A colour plot is presented in Figure \ref{fig:temp}(a) showing the final GS fidelity for the 9-spin system with varying $\Delta J_\text{xx}$ and $t_a$. We first note that we observe a GS fidelity close to unity for a range of $t_a$ and $\Delta J_\text{xx}$ values. Notably (as can also be seen from Figure \ref{fig:dynamics} discussed in the previous section), the final GS fidelity remains unity as $t_a$ is increased if $\Delta J_\text{xx} = 0$. However, for finite $|\Delta J_\text{xx}|$, we observe a drop-off in fidelity with increasing $t_a$.

To see this more clearly, we plot the final GS fidelity against $t_a$ for different values of $\Delta J_\text{xx}$ in Figure \ref{fig:temp}(b). The slices of Figure \ref{fig:temp}(a) that the curves in this plot correspond to are indicated with white dashed lines. We find that this decrease in fidelity is well fit by an exponential decay, with the decay rate increasing with $\Delta J_\text{xx}$. That there is a drop-off in fidelity for very short anneal times for all $\Delta J_\text{xx}$ is a result of the dynamics being too fast for the system to follow the instantaneous states and so the system becomes distributed over the state space. In the limit of $t_a = 0$, where the system has no time to evolve at all, the system ends in the equal superposition state and so the final GS fidelity is $1/2^n$. 

These results tell us that if $J_\text{xx}$ is selected with perfect precision, the enhancement to the final GS fidelity is robust to changes in the total anneal time, given that $t_a$ is sufficiently long for negligible amplitude to leak into other eigenstates. However, as we increase the $\Delta J_\text{xx}$, a greater demand is placed on the value chosen for $t_a$. If $t_a$ is too large, the fidelity will have decayed significantly from its maximum value and if $t_a$ is too small we will be in the regime where the dynamics are too fast which will also reduce the fidelity. As such, we can say that the less precision we have in $J_\text{xx}$, the more precision we need in our selection of $t_a$.

We can also consider how the precision in $t_a$ effects robustness to imprecision in $J_\text{xx}$. Looking at $t_a = 6\mu$s in Figure \ref{fig:temp}(b), we can see that the by choosing an unfavourable anneal time, the final GS fidelity has a strong dependence on $\Delta J_\text{xx}$ - with its value ranging from unity to zero depending on the catalyst strength chosen. Whereas for $t_a \approx 0.5\mu$s, the fidelity does not drop below $0.7$ for the largest $\Delta J_\text{xx}$. As such, we can say that the greater our precision in choosing a favourable anneal time, the more robust the final GS fidelity becomes to imprecision in $J_\text{xx}$. 

To see this more clearly we plot the final GS fidelity with varying $\Delta J_\text{xx}$ for three different values of $t_a$ in Figure \ref{fig:temp}(c). The slices of Figure \ref{fig:temp}(a) that these curves correspond to are again indicated with white dashed lines - with the exception of the $t_a = 10 \mu$s curve which lies outside the range of  Figure \ref{fig:temp}(a). For $t_a = 3$, $5$ and $10\mu$s, a fidelity of unity is obtained for $\Delta J_\text{xx} = 0$. The full width half maxima (FWHM) of the curves decreases as $t_a$ is increased, indicating a greater demand on the precision needed in $J_\text{xx}$. While the curves continue to increase in width for shorter anneal times, the peaks begin to decrease, reflecting the system delocalising across the state space as a result of the dynamics being too fast. 

In this section we have observed a trade-off between the precision needed in the anneal time, $t_a$, and the catalyst strength, $J_\text{xx}$, in order to preserve enhancement to the final GS fidelity that can be achieved through the creation of a diabatic path. While these results are for a very specific system and catalyst, we will make the case that these findings are likely to apply to other settings where there is some parameter that tunes the gap sizes in the annealing spectrum.

We briefly note that, for the $5$-spin system, oscillations in the final GS fidelity with increasing $t_a$ were observed. These oscillations have little impact on the fidelity in comparison to the decays and are also not present for the larger system sizes. We therefore do not take the time to discuss them here. We suspect, however, that these oscillations are the result of the same interference discussed in \cite{Munoz_Bauza_2019}.

\subsection{ Landau-Zener transitions \label{sec:LZ} }

\begin{figure}[]
    \centering
    \includegraphics[scale=0.27]{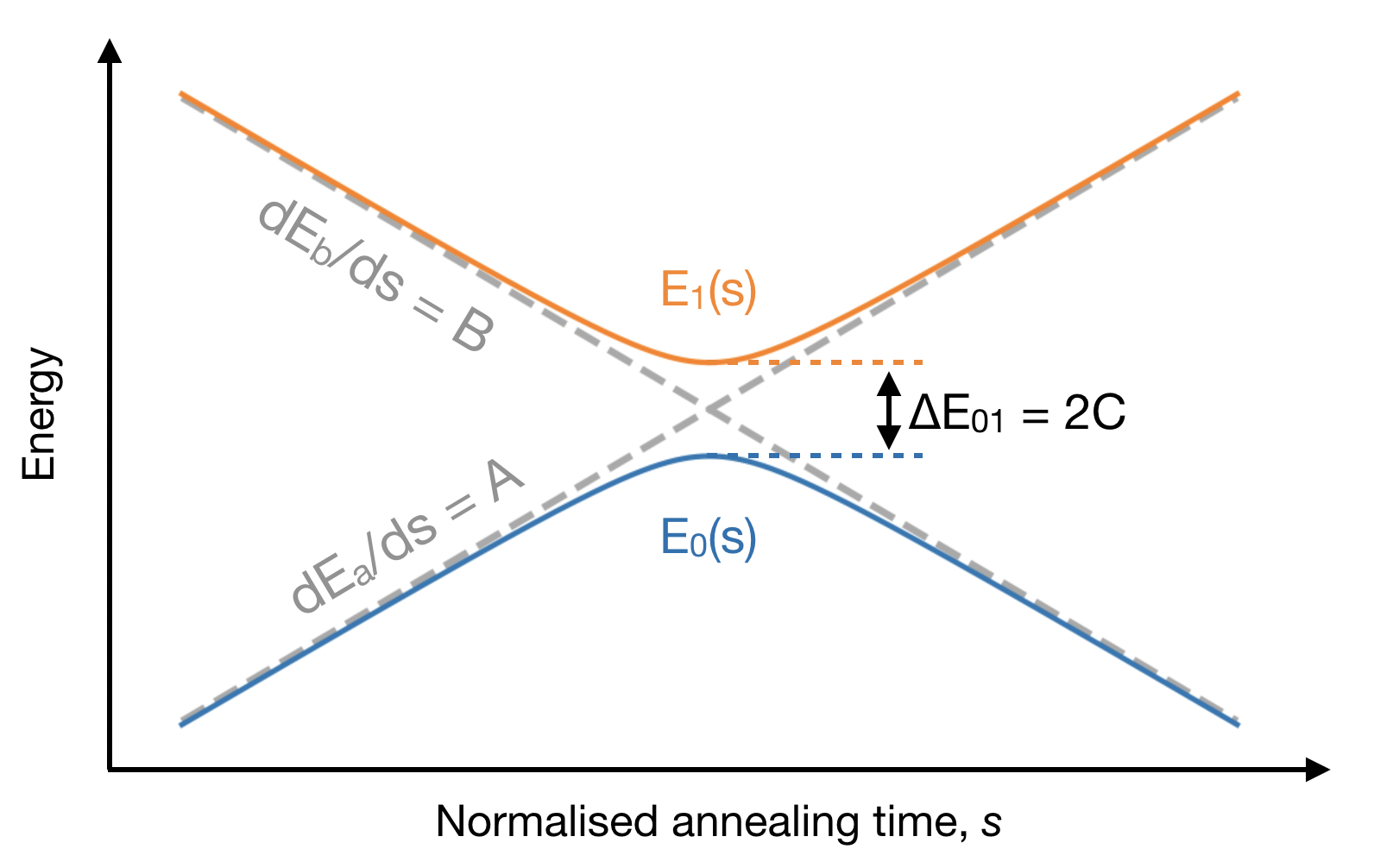}
    \caption{ Cartoon of an avoided level crossing, illustrating the key components that enter the Landau-Zener formula. }
    \label{fig:cartoon}
\end{figure}

We now discuss the physics behind the observations that our numerical work has revealed. We begin by making some observations on the dynamics of the individual anneals to allow us to build a qualitative picture of the mechanisms behind our results. These observations will also serve as assumptions when, in the second part of this section, we use a Landau-Zener (LZ) model to lend more rigour to this description. The predictions from the LZ-model are in excellent agreement with the numerically obtained fidelities.

The first observation we make is that, so long as we restrict ourselves sufficiently long anneal times (for a given system size), no amplitude exchange happens outside of the locations of the two gap minima. As such, we only need to consider the dynamics at the location of the perturbative crossing and the tunable gap minimum created by the catalyst. In addition to this, we need only concern ourselves with the subspace spanned by the instantaneous ground and first excited states. 

Next, we note that size of the gap minima at the perturbative crossings are such that the amplitude is fully exchanged at this point - \textit{i.e.} any amplitude in the first excited state will be transferred to the ground state and \textit{vice versa}. For this to not be the case, we would need to increase the annealing time by several orders of magnitude - as indicated by the large anneal times needed to reach the adiabatic limit. 

Putting these two observations together, we can expect the final GS fidelity to be precisely the amplitude transferred to the first excited state at the tunable gap minimum created by the catalyst. Thus, when this gap approaches zero, (\textit{i.e.} we have a small enough $|\Delta J_\text{xx}|$), we can expect the fidelity to be unity for a range of annealing times - since all the amplitude will be transferred out of the ground state. However, as we increase the gap size by increasing $|\Delta J_\text{xx}|$, the same run-times will result in less amplitude being transferred to the 1ES, and so a lower final GS fidelity - as observed in Figure \ref{fig:temp}(c). Correspondingly, keeping $\Delta J_\text{xx}$ the same and increasing the run-time will result in a lower fidelity - as seen in Figure \ref{fig:temp}(b).

Let us now make this description more quantitative by using the Landau-Zener (LZ) formula to obtain an expression for the diabatic transition rate at the new gap minimum created by the catalyst. From the preceding observations, we can assume that the system is completely in the ground state prior to this gap minimum and that we can restrict ourselves to the subspace spanned by $\ket{E_0(s)}$ and $\ket{E_1(s)}$ - treating the system as a two level system. The LZ formula for the diabatic transition rate is
\begin{equation}
        P_D = \exp{ \left( -2\pi\frac{C^2}{|\frac{d}{dt}(E_a - E_b)|} \right) }
        \label{eq:PD}
\end{equation}
where $C$ is the off-diagonal element coupling the states of the two-level system and $E_{a,b}$ are two linear crossing energy levels between which a finite gap emerges due to level repulsion. Substituting in $t = s \times t_a$, we can rewrite this expression as 
\begin{equation}
    P_D = \exp{ \left(-2\pi\frac{C^2}{|A-B|}t_a\right) }.
        \label{eq:PD2}
\end{equation}
where $A = dE_a/ds$ and $B = dE_b/ds$. A cartoon illustrating this is presented in Figure \ref{fig:cartoon}. We have argued that the final GS fidelity will be equal to $P_D$ and so we expect the decay rate of this fidelity with $t_a$ to be
\begin{equation}
    \Gamma = 2\pi\frac{C^2}{|A-B|}.
    \label{eq:decayrate}
\end{equation}

To obtain expressions for $A$, $B$ and $C$ we introduce the 2-level Hamiltonian resulting from the following expansion around the location of the gap minimum, $s^*$, which we use to first order:
\begin{equation}
    H^\text{2-level}(s) = \Bar{E}(s^*) \mathbb{I} + 
    \begin{pmatrix} 
    A(s-s^*) & C \\
    C & B(s-s^*) 
    \end{pmatrix}
    + \mathcal{O}(s^2).
    \label{eq:expansion}
\end{equation}
Here, $s^*$ denotes the location of the gap minimum, $\mathbb{I}$ is the identity matrix and $\bar{E}(s)$ is the average energy of the instantaneous ground and first excited state, $\frac{1}{2}(E_0(s)+E_1(s))$. By obtaining and differentiating the energies, $E^\text{2-level}_{0,1}$, of this Hamiltonian, one arrives at the following expressions for $A$, $B$ and $C$:
\begin{equation}
    C = \frac{1}{2} \times \Delta E^\text{2-level}_{01},
    \label{eq:C}
\end{equation}
\begin{equation}
    A = E^\text{2-level}_{0,1}{'} + \sqrt{\frac{1}{2} \Delta E^\text{2-level}_{01} \times |E^\text{2-level}_{0,1}{''}|}\text{ ,}
    \label{eq:A}
\end{equation}
\begin{equation}
    B = E^\text{2-level}_{0,1}{'} - \sqrt{\frac{1}{2} \Delta E^\text{2-level}_{01} \times |E^\text{2-level}_{0,1}{''}|}\text{ .}
    \label{eq:B}
\end{equation}
where the energies and their derivatives are those at $s=s*$. A full description of how these expressions are obtained can be found in Appendix \ref{app:LZ}

Substituting Equations \ref{eq:C}-\ref{eq:B} into Equation \ref{eq:decayrate} one finds:
\begin{equation}
    \Gamma = \frac{\pi}{\sqrt{2}} \frac{\Delta E_{01}(s^*)^2}{\sqrt{\Delta E_{01}(s^*) \times |E_{0,1}''(s^*)|}}.
    \label{eq:decayrate_2}
\end{equation}
The 2-level energies and their derivatives can then be equated with the corresponding numerical results for the full Hamiltonian at the location of the gap minimum. \textit{i.e:} $\Delta E^\text{2-level}_{01} = \Delta E_{01}(s^*)$ and $E^\text{2-level}_{0,1}{''} = E_{0,1}''(s^*)$.  The values for $\Delta E_{01}(s^*)$ can be trivially extracted from the numerical data and the second derivatives, $E_{0,1}''(s^*)$, are computed by finite differences. 

For these second derivatives, we have a choice of using either $E_{0}''(s^*)$ or $E_{1}''(s^*)$. Equation \ref{eq:E''}, found in Appendix \ref{app:LZ}, suggests that the two values will have the same magnitude however this will only be true for an actual 2-level system. Figure \ref{fig:spectrum_properties}(b) shows the numerically obtained values for $|E_{0}''(s^*)|$ and $|E_{1}''(s^*)|$ with varying $\Delta J_\text{xx}$ for the 9-spin system. We see that the values obtained for $E_{0}''(s^*)$ are larger in magnitude than those for $E_{1}''(s^*)$ and will therefore result in a smaller prediction for the decay rate. 

The predictions for the decay rates corresponding to the 9-spin system are compared to the numerical results in Figure \ref{fig:temp}(b). We show the area enclosed by the decay predictions obtained using either $E_{0}''(s^*)$ or $E_{1}''(s^*)$ in grey. The decay rates obtained by averaging the two results are shown in dashed grey curves. We see that there is excellent agreement between these decays and the numerical results. That the numerical results don't follow the LZ predictions for $t_a < 0.5\mu$s is because the assumption that no amplitude exchange happens away from the two gap minima no longer holds.

\section{ Discussion \label{sec:discussion} }

\begin{figure}[]
    \centering
    \includegraphics[scale=0.28]{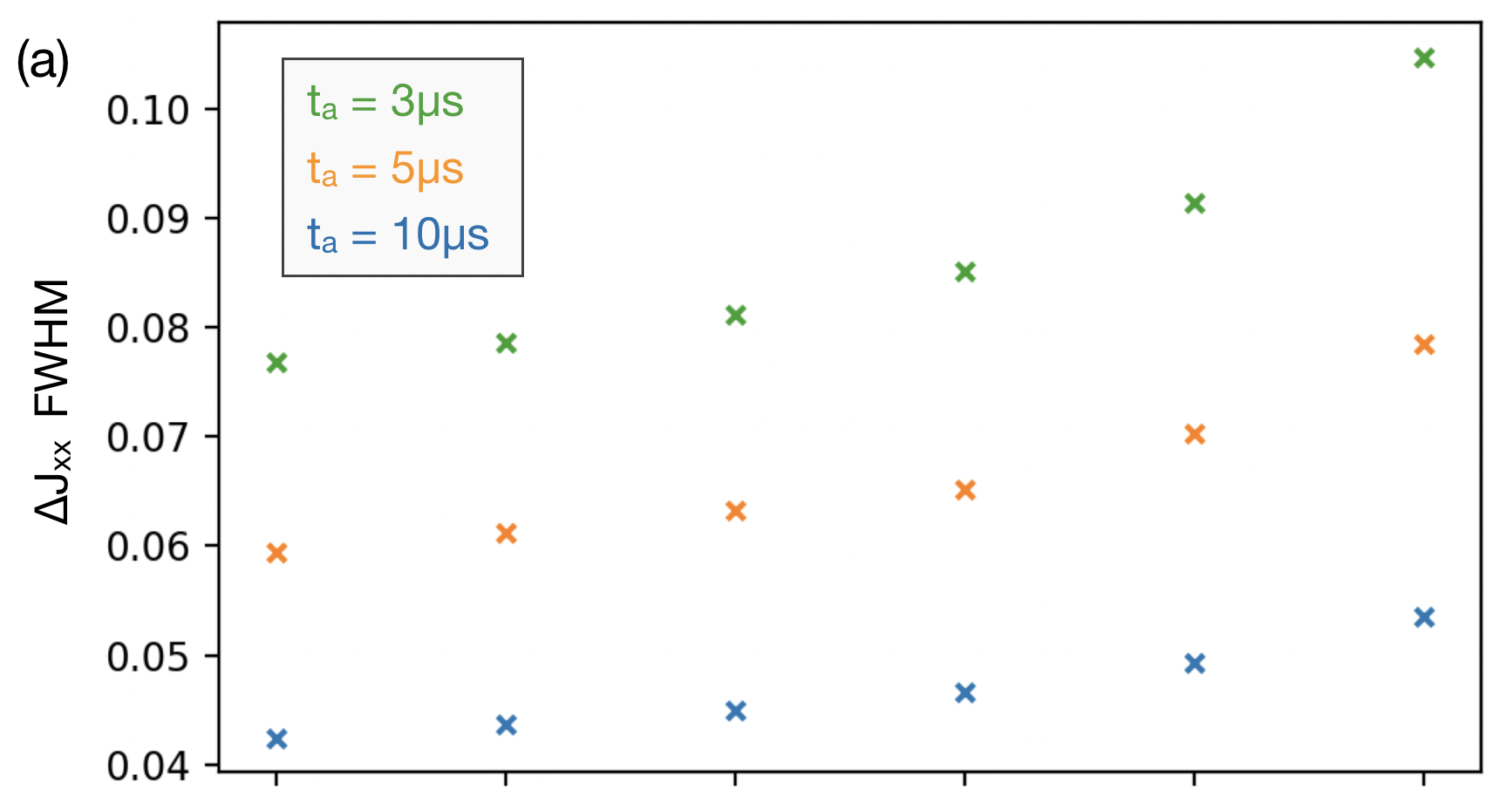}
    \includegraphics[scale=0.28]{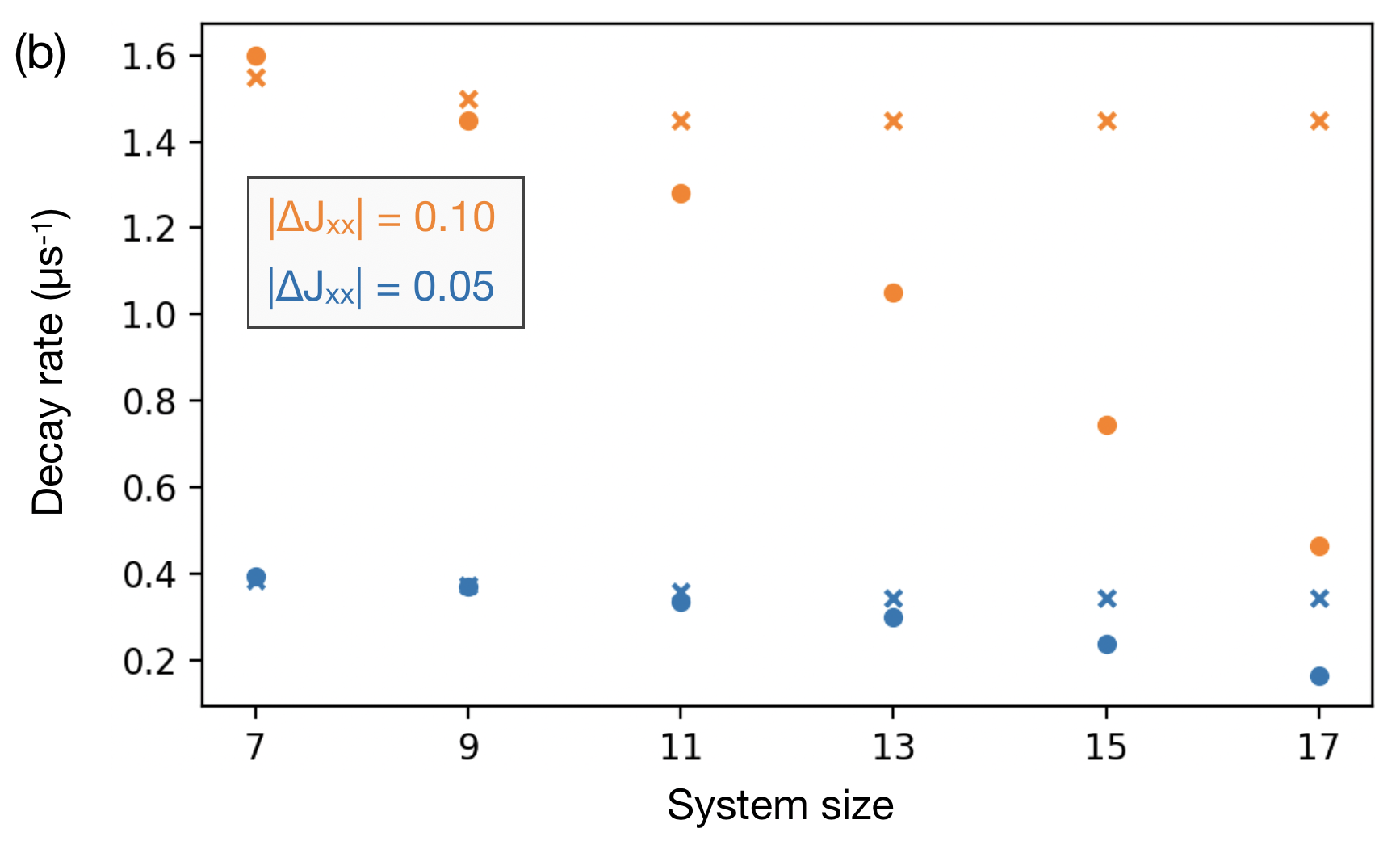}
    \caption{ Plot (a) shows the $\Delta J_\text{xx}$ FWHM of the final GS fidelity for three different anneal times. Plot (b) shows the decay rate of the final GS fidelity against system size when $|\Delta J_\text{xx}| = 0.05$ in blue and when $|\Delta J_\text{xx}| = 0.10$ in orange. Results for positive and negative $\Delta J_\text{xx}$ are plotted with dots and crosses respectively. The values in this figure are obtained from the numerical data. }
    \label{fig:scaling}
\end{figure}

\begin{figure}[]
    \centering
    \includegraphics[scale=0.36]{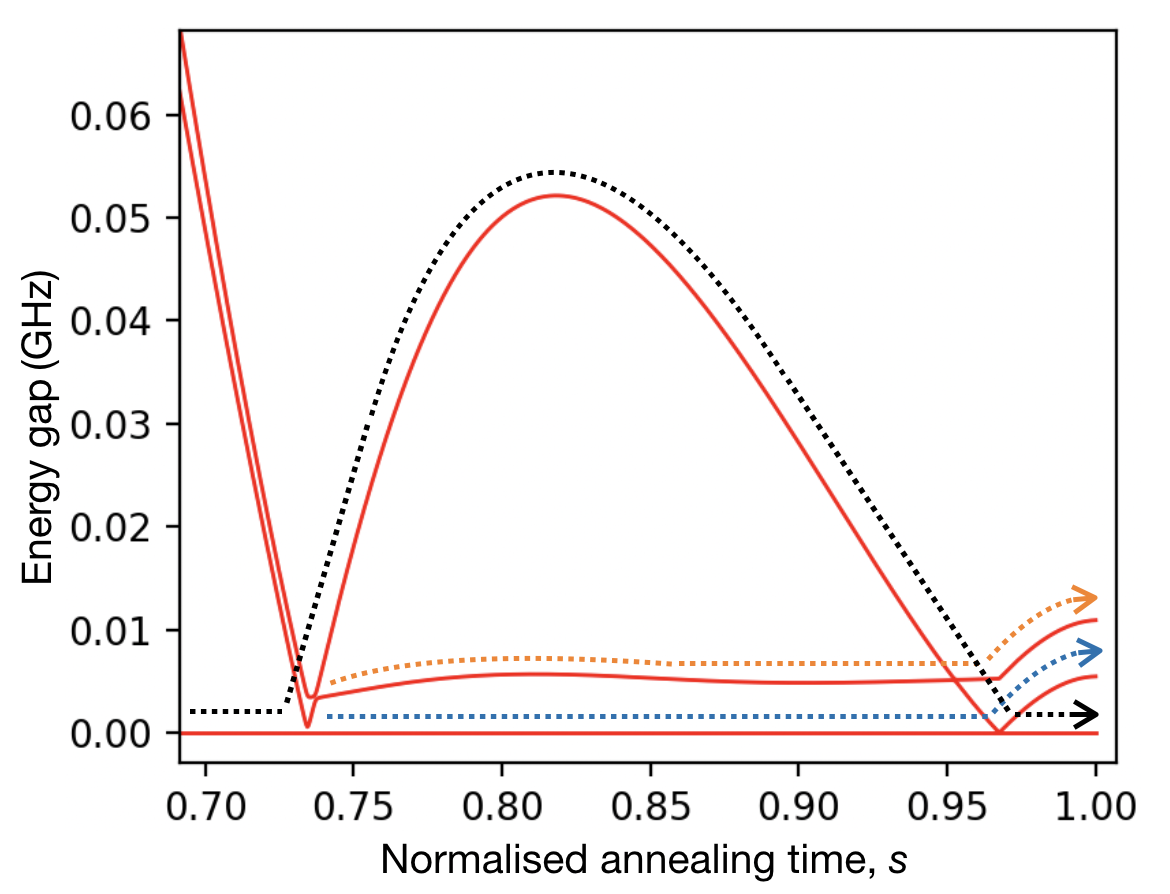}
    \caption{ Gap spectrum corresponding to a tripartite $8$-vertex graph where a catalyst has been introduced to create a diabatic path to the final GS. (Details of this graph and catalyst can be found in Appendix \ref{app:tripartite}.) The black dotted curve shows the diabatic path we want the system to take. The blue and orange dotted curves show potential amplitude losses from the system not fully transitioning at the new gap minima created by the catalyst. }
    \label{fig:tri-partite}
\end{figure}

In this study, we have used a toy model to investigate the robustness of the enhancement to the final GS fidelity by DQA in the closed system setting. The setting has a small gap minimum towards the end of the annealing spectrum at which the system transitions out of the GS if the anneal is run too quickly. Through the introduction of a carefully chosen catalyst Hamiltonian, an additional gap minimum can be introduced into the annealing spectrum - the size of which depends on the strength with which the catalyst is introduced. The tunability of the size of this additional gap minimum, as well as the simplicity of the diabatic path that it creates makes this setting an ideal testing ground for the robustness of DQA. 

We numerically determined that if $J_\text{xx}$ is selected such that the new gap minimum is suppressed to near zero, the final GS fidelity is robust to changes in the anneal time. That is, so long as the anneal is run slowly enough that the system does not delocalise across the state space, a GS fidelity of unity is obtained. If we move away from this value of $J_\text{xx}$, such that the gap minimum is larger, we see an exponential decrease in the final GS fidelity with $t_a$ - with the decay rate increasing with $|\Delta J_\text{xx}|$. Thus, as we decrease our precision in $J_\text{xx}$, we increase the demand on the precision in $t_a$ that is needed to maintain the enhancement to the final GS fidelity. Similarly, by selecting an appropriate anneal time, the effect that imprecision in $J_\text{xx}$ has on the final GS fidelity can be minimised. However, as we increase our anneal time away from this point, we reduce the robustness of the final GS fidelity to imprecision in $J_\text{xx}$. These results indicate a trade-off between the precision needed in the choice in anneal time, $t_a$, and catalyst strength, $J_\text{xx}$ - with greater precision in one resulting in greater robustness to imprecision in the other.

We attributed this to the fact that, when $J_\text{xx} = J^*_\text{xx}$, the gap minimum created by the catalyst is small enough that all the amplitude is transferred to the 1ES at this point for a range of $t_a$, such that it is all returned to the GS at the perturbative crossing. However, when $\Delta J_\text{xx} \neq 0$, the gap becomes larger and a sufficiently short anneal time must be chosen to ensure that the amplitude transferred to the 1ES is close to unity - but not so short that the dynamics are too fast for the system to follow them at all. As such, we expect these findings to apply to other settings where there is some parameter that controls the gap size at a point where we want a diabatic transition to occur - which is also the case in \cite{Choi2021} and \cite{Fry-Bouriaux2021}. 

In practice, what might lead to deviations from the optimal values of $J_\text{xx}$ (or other parameters that may be used to alter the spectrum,  \textit{e.g.} driver fields) and $t_a$? Firstly, we may have a lack of knowledge of what the optimal parameters actually are. Strategies for manipulating annealing spectra in the ways in which we are discussing are still very much in their infancy and so there is not a great deal to be said regarding to what precision we may expect to know these parameters. In the system we have studied in this work, we see that the optimal $J_\text{xx}$ value decreases with system size and appears as though it may be tending to a constant - Figure \ref{fig:with_catalyst}(c). However, we have examined a very specific system and we have not yet investigated how a suitable $J_\text{xx}$ may be estimated for an arbitrary system. For the catalyst examined in \cite{Choi2021}, the author proposes a method for obtaining an estimate for a range in which an appropriate $J_\text{xx}$ value can be found. In \cite{Fry-Bouriaux2021}, the driver field on a single target qubit is reduced to produce the new gap minimum. Here, the value of the parameter that minimises the gap is known to be zero and so there is no uncertainty in the optimal value of $J_\text{xx}$ that will make the final GS fidelity robust to changes in $t_a$. 

However, even if we can expect to know the optimal value of these parameters, there will also be hardware limitations to achievable precision. For instance in D-Wave quantum annealers, which consist of superconducting flux qubits, the precision of the local fields and coupling strengths is limited by, among other things, quantisation errors, flux noise and the fact that the qubits will not be manufactured to be perfectly identical. This can result in errors on the order of $0.01$ to the field and coupling strengths.

For any setting where the spectrum is being manipulated in the ways in which we have discussed, it will be important to understand the sensitivity of the spectrum to changes in the relevant parameter(s). And then, crucially, how this relates to the precision with which we expect to know the optimal parameter settings as well as the precision which the hardware can provide. This will indicate how precisely the anneal time will need to be chosen.

So far, we have discussed what we consider to be some general conclusions that can be drawn from this study and what they imply for the successful implementation of strategies where the annealing spectrum is manipulated to facilitate DQA. We have predominantly used the problem setting described in section \ref{app:prob_instances} as a toy model to investigate the robustness of DQA protocols in a stripped back setting. However, our results indicate that this catalyst may be useful in manipulating the annealing spectrum to facilitate DQA. In addition to our findings in \cite{feinstein2023effects}, we have also begun investigating the effect of this catalyst in more complex problems - with our initial results suggesting that spectra amenable to diabatic annealing can be produced in these more general settings as well. As such, we now turn to what specific conclusions can be drawn regarding the catalyst that we have used in this work. 

Specifically, we consider how the precision needed in $J_\text{xx}$ and $t_a$ scales with system size for our particular setting. We characterise the precision needed in $J_\text{xx}$ with the FWHM of the curves shown in Figure \ref{fig:temp}(c) and the precision needed in $t_a$ by the decay rate of the fidelities. We plot these values for increasing system size in Figures \ref{fig:scaling}(a) and (b) respectively. 

In Figure \ref{fig:scaling}(a) we see that the FWHM increases with the system size for the three anneal times plotted, indicating that less precision is needed as the system size increases. Figure \ref{fig:scaling}(b) shows the decay rates when $\Delta J_\text{xx} = \pm 0.05$ and $\pm 0.10$. For the positive $\Delta J_\text{xx}$ values we see a decrease in the decay rate with increasing system size - again indicating that less precision is needed for the larger systems. For the negative $\Delta J_\text{xx}$ values there is also some reduction in the decay rate however the values quickly plateau. The reduction in the required precision can be attributed to a decrease in the rate at which the gap size increases with $J_\text{xx}$ - as seen in \ref{fig:spectrum_properties}(a).  

We note that while this is a promising result for the catalyst that we study here and in \cite{feinstein2023effects}, the situation is not so simple. That the rate of gap increase with $|\Delta J_\text{xx}|$ becomes smaller for larger system sizes goes hand in hand with the fact that the spectral gaps generally decrease with increasing system size. This means that larger anneal times will be needed to prevent amplitude leaking between the instantaneous states away from the two gap minima. For the larger system sizes that we have examined, we already observe a suppression to the peak fidelity reached for the annealing times that we have looked at ($\leq 20 \mu$s). As already discussed, an increase in $t_a$ results in an increase to the precision needed in $J_\text{xx}$. It will be important therefore to study the interplay between this and the data presented in Figure \ref{fig:scaling} in order to understand how the precision needed would scale in practice. 

We have chosen to study a very simple setting with a straightforward diabatic path to the final GS in order to explore the precision required to maintain the enhancement to the final GS fidelity by DQA. In practice, the manipulations to the annealing spectrum may be such that the diabatic path may involve many transitions between energy levels. For instance, in our initial investigations into applying the catalyst we have examined in this work to more general systems, we have found that a diabatic route to the GS that involves more energy levels can be created. An example is shown in Figure \ref{fig:tri-partite}. The setting that produces this spectrum is outlined in Appendix \ref{app:tripartite}. The two ACs towards the end of the spectrum are present in the catalyst free case and the gap minimima between the ground, first and second excited states at around $s=0.74$ are created by the catalyst. As with the toy example we have studied here, the gaps at the ACs at the end of the spectrum are small enough that, unless very long anneal times are chosen, there will be a complete exchange in amplitude at these points. We therefore expect the effect of increasing the sizes of the gap minima created by the catalyst to be much the same as our findings here - with the effects compounding at each gap minimum since any amplitude not transferred to the next energy level will not be returned to the GS. This is illustrated in Figure \ref{fig:tri-partite} with coloured arrows.

\section{ Conclusion \label{sec:conclusion} }

We have used a simple system with a tunable gap minimum in the annealing spectrum to investigate the robustness of DQA to changes in the anneal time and parameters controlling the spectrum. We determined that there was a trade-off between precision needed in $J_\text{xx}$ and $t_a$ to maintain the enhancement to the final GS fidelity provided by the catalyst. We understood this trade-off through the theory of LZ transitions, suggesting that these findings should apply to other settings where there is some parameter that controls the gap size at a point where we want a diabatic transition to occur - as is the case in other works where the spectrum is manipulated to be amenable to DQA \cite{Choi2021,Fry-Bouriaux2021}.

Crucially, it will be important to understand how sensitive of the spectrum is to changes in the parameter(s) being used to alter it and how this relates to the precision with which we expect to know the optimal parameter settings as well the hardware limitations. This will indicate how precisely the anneal time will need to be chosen.

For the specific catalyst examined here, our initial results for how the precision needed scales with the system size suggested that the demands on the precision needed in $J_\text{xx}$ and $t_a$ may decrease with system size. However, as we have noted, the increase in the anneal time required for the evolution to be approximately adiabatic away from the two gap minima means that how the precision needed scales in practice may be much less favourable. 

\begin{acknowledgments}
The authors would like to thank Robert Banks for insightful discussions and Tameem Albash for his help in implementing the Hilbert space reductions. This work is supported by EPSRC, grant references EP/S021582/1 and EP/T001062/1.
\end{acknowledgments}

\bibliography{refs.bib}

\appendix

\section{ Perturbative Crossings \label{app:PCs} }

The formation of perturbative crossings can be understood by introducing $H_d$ as a perturbation to $H_p$:
\begin{equation}
    H(\lambda) = H_p + \lambda H_d.
    \label{eq:H-pert}
\end{equation}
The perturbed energies up to second order corrections are then given by
\begin{multline}
    E_a(\lambda) = E_a + \lambda \braket{E_a|H_d|E_a} + \lambda^2 \sum \limits_{c \neq a} \frac{|\braket{E_c|H_d|E_a}|^2}{E_a - E_c} \\ = E_a + \lambda^2 \sum \limits_{c \in N_\text{X}(a)} \frac{1}{E_a - E_c}
    \label{eq:general-pert}
\end{multline}
where we have used $N_\text{X}(a)$ to denote the neighbourhood of $a$ - \textit{i.e}: the set of states to which $a$ is coupled by the driver. To get from the first to the second line of equation \ref{eq:general-pert}, we have used the fact that $|\braket{E_b|H_d|E_a}|=1$ for $b \in N_\text{X}(a)$ and $0$ otherwise. The first order term vanishes since $H_d$ cannot couple a problem state to itself.  

A crossing occurs between two problem states $a$ and $b$ ($b > a$) if $E_b(\lambda) < E_a(\lambda)$ for some $\lambda$. If this $\lambda$ is small enough that perturbation theory remains valid, this indicates the formation of an avoided level crossing (AC) towards the end of the instantaneous gap spectrum - where the finite gap size is a result of tunnelling between the two states. In general, the crossings of interest are those between the lowest energy instantaneous states and, in particular, those involving the instantaneous ground state. We therefore restrict ourselves to considering problem states with $E_a = E_0 + \delta$ where $\delta$ is small compared to the energy gap between the problem ground state and the rest of the spectrum. We refer to this set of problem states as $L_\delta$. 

If all states in $L_\delta$ correspond to local optima, then $E_b > E_a$ for all $a \in L_\delta$ and $b \in N_\text{X}(a)$. As such, each term in the sum from equation \ref{eq:general-pert} will be negative for all states $a \in L_\delta$. For these states, we can subtly rewrite equation \ref{eq:general-pert} for clarity as 
\begin{equation}
    E_a(\lambda) = E_a - \lambda^2 \sum \limits_{c \in N_\text{X}(a)} \frac{1}{|E_c - E_a|}.
    \label{eq:low-energy-perts}
\end{equation}
Taking of the absolute value in the denominator is redundant since $E_c-E_a$ will always be positive in this context however we include it to make the sign of these terms explicit. Since the contributions to the sum are all cumulative, it is clear that a state $a \in L_\delta$ will receive a greater negative perturbation if it has more states in its neighbourhood that are close in energy to it. And since $L_\delta$ contains the very lowest energy states, the neighbours that provide the greatest contributions to the sum in equation \ref{eq:low-energy-perts} will be the lowest energy neighbours. 

Thus, for a crossing to occur between two states $a$ and $b \in L_\delta$ ($b>a$), $N_\text{X}(b)$ must contain more low energy states than $N_\text{X}(a)$. Explicitly, there must be a positive value of $\lambda$ for which
\begin{multline}
    E_a(\lambda) = E_b(\lambda), \\
     E_a - \lambda^2 \sum \limits_{c \in N_\text{X}(a)} \frac{1}{|E_c - E_a|} =  E_b - \lambda^2 \sum \limits_{c \in N_\text{X}(b)} \frac{1}{|E_c - E_b|}, \\
     \Delta E_{ab} = \lambda^2 \left( \sum \limits_{c \in N_\text{X}(b)} \frac{1}{|E_c - E_b|} - \sum \limits_{c \in N_\text{X}(a)} \frac{1}{|E_c - E_a|} \right)
\end{multline}
where we have used $\Delta E_{ab}$ to denote the energy difference $E_b - E_a$. In order for the crossing to occur at small $\lambda$, such that perturbation theory remains valid, the difference in the energies of the neighbourhoods around $a$ and $b$ must be large enough compared to the unperturbed energy gap $\Delta E_{ab} = \mathcal{O}(\delta)$.

The size of the gap minimum that forms at the resultant AC is proportional to the overlap between the two perturbed states at the point of the crossing. The perturbed states can be written as 
\begin{equation}
    \ket{E_a(\lambda)} = \sum \limits_{c = 0}^{2^n -1} c_{ac}(\lambda) \ket{E_c},
\end{equation}
where the $c_{ac}(\lambda)$ values are the components of the perturbed vector $\ket{E_a(\lambda)}$ in the computational basis - or equivalently, the overlaps of $\ket{E_a(\lambda)}$ with the problem eigenstates. From this we can write
\begin{equation}
    \Delta E_{ab}(\lambda^*) \propto \braket{E_a(\lambda^*)|E_b(\lambda^*)} =  \sum \limits_{c = 0}^{2^n -1} c_{ac}(\lambda^*)c_{bc}(\lambda^*)
    \label{eq:gap-prop}
\end{equation}
where we have used $\lambda^*$ to denote the location of the crossing. Perturbation theory tells us that, for any perturbed problem state $a$, the magnitude of $c_{ac}(\lambda^*)$ will be exponentially small in the Hamming distance between $\ket{E_a}$ and $\ket{E_c}$ - since $\ket{E_c}$ will only enter the perturbative expansion on the order of this Hamming distance. Because each term in equation \ref{eq:gap-prop} will depend on both the Hamming distance between each state, $a$ and $b$, and the state $c$, we can expect $\Delta E_{ab}(\lambda^*)$ to decrease exponentially with the Hamming distance between $\ket{E_a}$ and $\ket{E_b}$. This Hamming distance will, to some extent, be instance specific - in that it depends on how well separated the problem's local optima are. However, we can generally expect this Hamming distance to grow linearly with the number of spins in the system resulting in a gap minimum that closes exponentially with the problem size.

\section{ Problem instances \label{app:prob_instances} }

The MWIS problem Hamiltonian is given by 
\begin{equation}
\label{eq:Prob-Hamiltonian-app}
    H_p = \sum \limits_{i \in \{\textrm{\scriptsize vertices}\}} (c_i J_{zz} - 2w_i)\sigma^z_i + \sum \limits_{(i,j) \in \{\textrm{\scriptsize edges}\}} J_{zz} \sigma^z_i \sigma^z_j
\end{equation}
where $c_i$ is the degree and $w_i$ the weight on vertex $i$. $J_{zz}$ is the edge penalty. As discussed in Section \ref{sec:nocat}, our problem instances consist of two disconnected subgraphs, $G_0$ and $G_1$, with edges connecting every vertex in $G_0$ to every vertex in $G_1$. We allocate a total weight of $1$ to $G_1$ and a total weight of $1 + \delta W$ to $G_0$ such that we have a highly competitive local optimum. This weight is split evenly between the vertices in the subgraphs. Equation \ref{eq:Prob-Hamiltonian-app} can then be written as
\begin{multline}
\label{eq:specific-problem}
    H_p = (n_1 J_{zz} - \frac{2(1+\delta W)}{n_0}) \sum \limits_{i \in G_0} \sigma^z_i \\ + (n_0 J_{zz} - \frac{2}{n_1})\sum \limits_{i \in G_1} \sigma^z_i \\ + J_{zz} \sum \limits_{i \in G_0} \sum \limits_{j \in G_1} \sigma^z_i \sigma^z_j,
\end{multline}
where $n_0$ and $n_1$ are the number of vertices in $G_0$ and $G_1$ respectively.

To keep the energy scale of $H_p$ relative to $H_d$ consistent for different problem sizes, we introduce a normalisation factor, $K$. For a set of un-normalised parameter values $(\delta W', J'_{zz})$ we first calculate the un-normalised vertex weights as $w'_i = (1+\delta W' )/n_0$ for $i \in G_0$ and $w'_i = 1/n_1$ for $i \in G_1$ - giving us our set of un-normalised parameters $(\{w'_i\}, J'_{zz})$. The normalised parameters are then obtained as
\begin{equation*}
    w_i = E_\textrm{\scriptsize scale} \times K \times w'_i,
\end{equation*}
\begin{equation*}
    J_{zz} = E_\textrm{\scriptsize scale} \times K \times J'_{zz}
\end{equation*}
where $E_\textrm{\scriptsize scale}$ sets the energy scale in relation to the driver and $K$ is a normalisation factor given by
\begin{equation*}
    K  = \frac{n_0 + n_1}{4(n_0 \times n_1 \times J'_{zz} - 1)}.
\end{equation*}
This normalisation factor has been obtained by using equation \ref{eq:specific-problem} to calculate the difference in energy between the ground and highest excited state.

The un-normalised problem parameters used in this work are $\delta W' = 0.01$ and $J'_{zz} = 5.33$ for all system sizes. The specificity of these parameters relates to the way in which they were chosen in our previous work \cite{feinstein2023effects}. They need not be this specific for the resultant annealing spectra to have the properties key to the discussions in this work. We do note however that for significantly (over an order of magnitude) larger values of $\delta W'$ and $J'_{zz}$, alternative effects from the catalyst Hamiltonian are observed \cite{feinstein2023effects}.

\section{ Hilbert space reduction \label{app:hilbert-space} }

In order to make our simulations more tractable, we use the symmetries in the Hamiltonian to reduce the size of the Hilbert space - utilising the same approach used in \cite{Albash2019}. 

In the catalyst free case, the Hamiltonian, $H(s)$ is invariant under permutation of qubits within $G_0$ and $G_1$. Introducing the catalyst effectively splits off two qubits from $G_1$. The Hamiltonian is now invariant under permutations within the subgraphs $G_0$, $G_{1'}$ and $G_c$ where $G_{1'}$ consists of all the qubits in $G_1$ not acted on by the catalyst and $G_c$ consists only of the two qubits with an XX-coupling between them. The system is initialised in the equal superposition state meaning the evolution starts in the subspace spanned by the states that are symmetric under these permutations. In the closed system setting, the evolution is confined to this subspace,  meaning that we can drastically reduce the size of the Hilbert space we need to consider. 

A natural basis for this subspace can be constructed from Dicke states associated with each of the subgraphs $G_0$, $G_{1'}$ and $G_c$. For each subgraph, $G_a$, we can write the total spin operator, 
\begin{equation*}
    S^z_a = \frac{1}{2} \sum \limits_{i \in G_a} \sigma^z_i,
\end{equation*}
of which the Dicke states, $\ket{s_a,m_a}$, are eigenstates. Their eigenvalues are $m_a$ which run from $-s_a$ to $s_a$ in integer steps.

In order to construct $H(s)$ out of the total spin operators $S^z_a$ and 
\begin{equation*}
    S^x_a = \frac{1}{2} \sum \limits_{i \in G_a} \sigma^x_i,
\end{equation*}
we explicitly calculate the matrix elements
\begin{equation*}
    \braket{s_a,m_a|S^z_a|s_a,m_a'} = \delta_{m_a,m_a'} \times m_a
\end{equation*}
and
\begin{multline*}
    \braket{s_a,m_a|S^x_a|s_a,m_a'} = \delta_{m_a,m_a'-1} \times \\ \frac{1}{2} \sqrt{s_a(s_a + 1) - m_a(m_a + 1)}.
\end{multline*}
$H(s)$ can then be constructed as follows:
\begin{multline}
    H(s) = (1-s) \times 2 \times (S^x_0 + S^x_{1'} + S^x_c) \\ 
    + s(1-s) \times 2 \times J_\text{xx} (S^x_c)^2 \\ 
    + s \times 2 \times [ h_0 S^z_0 + h_1(S^z_{1'} + S^z_c)
    \\ + 2 J_{zz} S^z_0 (S^z_{1'} + S^z_c ) ]
    \label{eq:reduced_H}
\end{multline}
where $h_{0/1}$ are the local field strength on the  spins as calculated in Appendix \ref{app:prob_instances}. We have dropped some identity terms that would make this Hamiltonian fully equivalent to the Hamiltonian defined by equations \ref{eq:H(s)}, \ref{eq:Hd}, \ref{eq:Problem-Hamiltonian} and \ref{eq:Hc}. However, this only causes a shift in the absolute values of the energies and we are concerned with the spectral gaps - which are unchanged. 

The states that span the symmetric subspace to which the evolution is restricted (and the eigenstates of equation \ref{eq:reduced_H}) are products of the Dicke states associated with the three symmetries:
\begin{equation}
    \ket{m_0,m_{1'},m_c} = \ket{\frac{n_0}{2},m_0} \otimes \ket{\frac{n_1}{2} - 1,m_{1'}} \otimes\ket{1,m_c}.
\end{equation}
The total number of states is $3 \times (n_0 + 1) \times (n_1 - 1) = 3(n-1)(n+1)/4$ where the factor of $3$ comes from the the three values that $m_c$ can take. Thus, the size of the Hilbert space now only scales polynomially with the system size.

\section{ Extracting the Landau-Zener parameters \label{app:LZ} }

Here we describe how we obtain the expressions for the LZ parameters in equations \ref{eq:C}-\ref{eq:B}. We begin with the 2-level Hamiltonian 
\begin{equation}
    H^\text{2-level}(s) = \Bar{E}(s^*) \mathbb{I} + 
    \begin{pmatrix} 
    A(s-s^*) & C \\
    C & B(s-s^*) 
    \end{pmatrix}
    + \mathcal{O}(s^2).
    \label{eq:expansion_app}
\end{equation}
where $s^*$ denotes the location of the gap minimum, $\mathbb{I}$ is the identity matrix and $\bar{E}(s)$ is the average energy of the instantaneous ground and first excited state, $\frac{1}{2}(E_0(s)+E_1(s))$. The only effect of the $\Bar{E}(s^*) \mathbb{I}$ term is to introduce a global phase to the time-dependent unitary. Since this does not affect the dynamics of the system, we may discard it in our analysis. Subtracting  $\Bar{E}(s^*) \mathbb{I}$ from equation \ref{eq:expansion_app}, we obtain
\begin{equation*}
    \Bar{H}^\text{2-level}(s) =
    \begin{pmatrix} 
    A(s-s^*) & C \\
    C & B(s-s^*) 
    \end{pmatrix}
    + \mathcal{O}(s^2)
\end{equation*}
where the energy levels are now centred on zero. This will simplify the following calculations. We can further simplify things by centering the gap minimum on zero with the change of variables $X=s-s^*$ such that we now have
\begin{equation}
    \Bar{H}^\text{2-level}(X) =
    \begin{pmatrix} 
    AX & C \\
    C & BX 
    \end{pmatrix}
    + \mathcal{O}(s^2).
    \label{eq:centred_expansion}
\end{equation}
The energies of the two eigenstates of this Hamiltonian are
\begin{equation}
    E^\text{2-level}_{0,1}(X) = \frac{(A+B)X \pm \sqrt{(A-B)^2 X^2 + 4C^2}}{2}
    \label{eq:LZ_E}
\end{equation}
By taking $X \rightarrow \pm \infty$, we find that the energies are linear in this limit and that their gradients are A and B - as desired. C can be read off Equation \ref{eq:centred_expansion} as the off-diagonal element. 

Setting $X=0$ in Equation \ref{eq:LZ_E} we get:
\begin{equation*}
    E^\text{2-level}_{0,1}(X=0) = \pm C,
\end{equation*}
\begin{equation}
    C = E^\text{2-level}_1 = -E^\text{2-level}_0 = \frac{1}{2} \times \Delta E^\text{2-level}_{01}
    \label{eq:C_app}
\end{equation}
where we have dropped the $X=0$ for readability. Differentiating equation \ref{eq:LZ_E} twice, with respect to $X$, and then setting $X=0$ one obtains
\begin{equation}
    \frac{dE^\text{2-level}_{0,1}}{dX}\bigg|_{X=0} = \frac{1}{2}(A+B)
    \label{eq:E'}
\end{equation} 
and 
\begin{equation}
    \frac{d^2E^\text{2-level}_{0,1}}{dX^2}\bigg|_{X=0} = \pm \frac{(A-B)^2}{4C}
    \label{eq:E''}.
\end{equation}
Using Equations \ref{eq:E'} and \ref{eq:E''}, we can obtain the following expressions for A and B: 
\begin{equation}
    A = E^\text{2-level}_{0,1}{'} + \sqrt{\frac{1}{2} \Delta E^\text{2-level}_{01} \times |E^\text{2-level}_{0,1}{''}|}
    \label{eq:A_app}
\end{equation}
\begin{equation}
    B = E^\text{2-level}_{0,1}{'} - \sqrt{\frac{1}{2} \Delta E^\text{2-level}_{01} \times |E^\text{2-level}_{0,1}{''}|}
    \label{eq:B_app}
\end{equation}
where we have switched to the prime derivative notation and, as in Equation \ref{eq:C_app}, we have dropped the $X=0$ for readability. 

\section{ Additional local optima \label{app:tripartite} }

Problems with the same properties as those described in \ref{sec:nocat} but with additional local optima can be generated by introducing one subgraph, $G_a$, for each local optimum to produce a complete $k$-partite graph. As with the bipartite graphs, each subgraph is given a total weight, $W_a$, that is divided evenly between its $n_a$ vertices. The local optima then correspond to picking all the vertices from one of the subgraphs. The energy spectra of the neighbourhoods of the corresponding problem states are determined by the size of the subgraphs, $n_a$. 

The gap spectrum presented in Figure \ref{fig:tri-partite} corresponds to a problem graph with three subgraphs. These have sizes $n_0 = 2$ and $n_1 = n_2 = 3$. The weights on these subgraphs are $W_0 = 1.010$, $W_1 = 1.005$ and $W_2 = 1.000$ and the edge penalty is kept as $5.33$. The catalyst Hamiltonian introduced to obtain the spectrum presented in Figure \ref{fig:tri-partite} consists of one XX-coupling in $G_1$ and another in $G_1$. Both of these are introduced with a strength of $J_\text{xx}=1.125$. As with the annealing spectra corresponding to the bipartite annealing spectra examined in this work, the gap minima present towards the end of the spectrum are also present in the catalyst free case. The catalyst is responsible for the gap minima observed between $s=0.70$ and $0.75$.

\end{document}